\documentstyle[emulateapj,apjfonts]{article}

\def\mkfigbox#1#2{
\hbox{ \epsfxsize=#2 \epsfbox{#1} \relax}
}

\def\kms{km s$^{-1}$} 
\def\etal{{\it et al.}}

\def\Sec{${}^{\prime\prime}$\llap{.}}

\def\etal{{\it et~al.\/}}

\def\kms{{km~s$^{-1}$}}
\def\kpc-1{{kpc$^{-1}$}}
\def\Mpc-1{{Mpc$^{-1}$}}
\def\s-1{{sec$^{-1}$}}
\def\pdeg2{{deg$^{-2}$}}

\def\h0{{H$_0$}}
\def\q0{{$q_0$}}

\def\rms{{\it rms\/}}

\def\etal{{\it et al.\/}}
\def\kms{\hbox{$\rm km\,s^{-1}$}}

\def\ltsima{$\scriptscriptstyle \; \buildrel < \over \sim \;$}
\def\simlt{\lower.3ex\hbox{\ltsima}}

\def\gtsima{$\scriptscriptstyle \; \buildrel > \over \sim \;$}
\def\simgt{\lower.3ex\hbox{\gtsima}}

\def\about{\raise.3ex\hbox{$\scriptscriptstyle \sim $}}

\def\Sec{\hbox{${}^{\prime\prime}$\llap{.}}}

\def\sqr#1#2{{\vcenter{\vbox{\hrule height.#2pt
        \hbox{\vrule width.#2pt height#1pt \kern#1pt
        \vrule width.#2pt}
        \hrule height.#2pt}}}}

\lefthead{Kelson \etal}
\righthead{$M/L_V$ Ratios in CL1358+62}

\begin{document}

\title{THE EVOLUTION OF EARLY-TYPE GALAXIES IN DISTANT CLUSTERS III.:
$M/L_V$ RATIOS IN THE $z=0.33$ CLUSTER CL1358+62
\altaffilmark{1,2}}

\author{Daniel D. Kelson\altaffilmark{3,4}, Garth D.
Illingworth\altaffilmark{4}, Pieter G. van Dokkum\altaffilmark{5,6}, and
Marijn Franx\altaffilmark{6}}

\altaffiltext{1}{Based on observations obtained at the W. M. Keck
Observatory, which is operated jointly by the California Institute of
Technology and the University of California.}

\altaffiltext{2}{Based on observations with the NASA/ESA {\it Hubble Space
Telescope}, obtained at the Space Telescope Science Institute, which is
operated by AURA, Inc., under NASA contract NAS 5--26555.}

\altaffiltext{3}{Department of Terrestrial Magnetism, Carnegie Institution
of Washington, 5241 Broad Branch Rd., NW, Washington, DC 20015}

\altaffiltext{4}{University of California Observatories / Lick Observatory,
Board of Studies in Astronomy and Astrophysics, University of California,
Santa Cruz, CA 95064}

\altaffiltext{5}{Kapteyn Astronomical Institute, P.O. Box 800, NL-9700 AV,
Groningen, The Netherlands}

\altaffiltext{6}{Leiden Observatory, P.O. Box 9513, NL-2300 RA, Leiden, The
Netherlands}

\begin{abstract}

Internal kinematics, length scales, and surface brightnesses, have
been determined for a large sample of 53 galaxies in the cluster
CL1358+62 at $z=0.33$ from Keck spectroscopy and {\it Hubble Space
Telescope\/} WFPC2 imaging over a $1.5h^{-1}\times 1.5h^{-1}$ Mpc$^2$
field of view. These data have been used to constrain the evolution of
early-type galaxies in the cluster environment.

We have constructed the fundamental plane using 30 E and S0 galaxies
and draw the following conclusions. The fundamental plane at $z=0.33$
has the following form: $r_e \propto \sigma^{1.31\pm 0.13} \langle
I\rangle_e^{-0.86\pm 0.10}$, similar to that found locally. The
1-$\sigma$ intrinsic scatter about this plane is $\pm 14\%$ in
$M/L_V$, comparable to that observed in Coma. We conclude that even at
intermediate redshifts, E and S0 galaxies are structurally mature and
homogeneous, like those observed in nearby clusters. The $M/L_V$
ratios of these early-type galaxies are offset from the Coma
fundamental plane by $\Delta \log M/L_V = -0.13\pm 0.03$ ($q_0=0.1$),
indicative of mild luminosity evolution in the stellar populations in
the CL1358+62 E and S0 galaxies. This level of evolution suggests a
luminosity-weighted formation epoch for the stars of $z > 1$. The
precise redshift depends on the initial mass function and on the
cosmology. The scatter about the fundamental plane is consistent with
that in the color-magnitude relation, indicating that the E/S0s have a
scatter in luminosity-weighted ages of $\simlt 15\%$.

We have also analyzed the $M/L_V$ ratios of galaxies of type S0/a and
later. These early-type spirals follow a different plane from the E
and S0 galaxies: $r_e \propto \sigma^{0.66\pm 0.29} \langle
I\rangle_e^{-0.64\pm 0.06}$, with a scatter that is twice as large as
the scatter for the E/S0s. The difference in the tilt between the
plane of the spirals and the plane of the E/S0s is shown to be due to
a systematic correlation of velocity dispersion with residual from the
plane of the early-type galaxies. These residuals also correlate with
the residuals from the color-magnitude relation. Thus for the spirals
in this cluster, as well as for the three E+A galaxies in the sample,
we see a systematic variation in the luminosity-weighted mean
properties of the stellar populations with central velocity
dispersion. If this is a relative age trend, then luminosity-weighted
age is positively correlated with $\sigma$, {\it i.e.\/}, more massive
spiral galaxies have older stars on average.

The residuals from the color-magnitude relation were used to correct
the surface brightnesses of the early-type spirals. After this
correction for age effects, these spirals fall on the fundamental
plane of E/S0s. We conclude that the early-type spirals may well
evolve onto the scaling relations of the old cluster members. After
correcting the spirals for the systematic trend with color, their
scatter about the fundamental plane does not decrease and remains
twice as large as that for the E/S0s. This large scatter should be
seen in a subsample of present-day cluster galaxies, if these spirals
evolve into contemporary S0s, unless some unknown process reduces
their scatter.

The colors and $M/L_V$ ratios imply that many cluster galaxies were
forming stars at least up until $z\sim 0.5$, but that this activity
was specific to the spiral population. The E+As will likely evolve
into low-mass present day S0 and Sa galaxies, but slightly bluer than
the present-day $(B-V)$ color-magnitude relation by $\about 0.08$ mag,
while fading to \about 1 mag below $L^*$.

\end{abstract}



\section{Introduction}

For many years, the evolution of galaxies has been estimated through
the use of galaxy counts (see the review by Koo \& Kron 1992), and,
more recently, using luminosity functions ({\it e.g.\/}, Lilly \etal\
1995). However, such data do not uniquely constrain the luminosity
evolution because of the unknown distribution of galaxy masses in a
given sample, since $M/L$ varies greatly with star-formation history.
Measurements of the mass scales of distant galaxies can remove this
ambiguity by allowing one to observe $M/L$ ratios and provide a
direct, unambiguous measure of the luminosity evolution. Reliable
estimates of the mass scales of galaxies at high redshifts can
routinely be made using the high spatial resolution imaging cameras on
the {Hubble Space Telescope\/} (HST), and efficient ground-based
spectrographs on 8-10m telescopes. HST data provide the requisite
length scales, and the spectrographs on 8-10m telescopes enable us to
accurately measure internal kinematics. Together, one can directly
measure $M/L$ ratios through the use of galaxy scaling relations in a
manner consistent with previous work on large samples of nearby
galaxies. These scaling relations, such as the Tully-Fisher relation
for spirals (\cite{tf77}), the Faber-Jackson relation for ellipticals
(\cite{fj76}), or the fundamental plane of early-type galaxies
(\cite{faber87,dd87}), are essentially relations between $M/L$ ratio
and galaxy mass. Using such relations, one can directly probe the
evolution of $M/L$ ratios for galaxies at high redshift, and set
powerful constraints on the evolution and formation histories of
normal galaxies.

The fundamental plane (FP) is an empirical relation between half-light
radius, $r_e$, surface brightness, $\langle I\rangle_e$, and central
velocity dispersion, $\sigma$ for early-type galaxies
(\cite{faber87,dd87}). The FP is a refinement of the Faber-Jackson
relation. Using a large sample of E and S0 galaxies in 11 clusters,
J\o{}rgensen, Franx, \& Kj\ae{}rgaard (1996) found, for 226 galaxies
in Gunn $r$ and 109 in Gunn $g$,
\begin{eqnarray}
r_e \propto \sigma^{1.24\pm 0.07} \langle I\rangle_e^{-0.82\pm 0.02},
&\qquad\hbox{(Gunn $r$)}
\nl
r_e \propto \sigma^{1.16\pm 0.10} \langle I\rangle_e^{-0.76\pm 0.04},
&\qquad\hbox{(Gunn $g$)}
\end{eqnarray}
Under the assumption of homology, the FP implies that $M/L$ ratio is a
tight function of galaxy structural parameters, such that, in Gunn
$r$,
\begin{equation}
M/L_{r} \propto M^{1/4} r_e^{-0.02}
\end{equation}
(\cite{faber87}). The fundamental plane is very thin (\cite{lucey}),
with an observed \rms\ scatter of $\pm 20\%$ in Coma in $V$-band $M/L$
ratio at a given $M$ (\cite{jfk93}).

By measuring a large sample of early-type galaxies at high redshift
({\it i.e.\/}, at large look-back time), one can use the evolution of
$M/L$ ratios and the evolution of the scatter in $M/L$ to infer the
star-formation histories of early-type galaxies. In this way, one
directly measures the star-formation history of the Universe in
high-density regions (where early-type galaxies are predominantly
found, {\it e.g.\/}, \cite{dress80}). Franx (1993, 1995) and van
Dokkum \& Franx (1996) used the MMT to obtain spectroscopy of several
early-type galaxies in the clusters A665 ($z=0.18$) and CL0024+16
($z=0.39$). They used structural parameters from HST imaging, and
velocity dispersions from the spectroscopy, to derive $M/L$ ratios for
early-type galaxies in those clusters. Due to the low internal scatter
in the fundamental plane, even these few observations sufficed to
measure mild evolution in the $M/L$ ratios as a function of redshift.
Several other authors have now measured elliptical galaxy scaling
relations at intermediate redshifts and have confirmed mild (passive)
stellar evolution out to moderate redshifts ($z\simlt 1$;
\cite{schade,ziegler,pahrethesis}).

Kelson \etal\ (1997) extended the fundamental plane measurements to
CL1358+62 ($z=0.33$) and MS2053--04 ($z=0.58$) and confirmed the
moderate $M/L$ evolution. Those authors also suggested that the
evolution in the scatter is likely to be quite low. Here, we present
the analysis of a larger sample of more than 50 galaxies in CL1358+62
at $z=0.33$. The spectroscopy and surface photometry are discussed by
Kelson \etal\ (1999a,b). Using these data, we now analyze the
fundamental plane in the cluster, with the goal of addressing the
following issues:

(1) What is the form of the fundamental plane at high redshift? Does
the slope evolve with redshift? What is the scatter, and does it
evolve with redshift? Is there simply a $M/L$ zero-point shift?

(2) Is there any {\it morphological\/} dependence in the fundamental
plane at high redshift, since large, nearby samples show no obvious
dependence or difference between S0s and ellipticals (\cite{jfk96})?

(3) Do $M/L_V$ ratios, for a given galaxy mass, correlate with known
stellar population indicators, such as $(B-V)$ color, or with
structural indicators, such as apparent ellipticity or profile shape?

(4) How do ``E+A'' galaxies relate to the typical early-type cluster
members? E+A galaxies have spectra which appear to be superpositions
of an old early-type stellar population, and a young 1-2 Gyr-old
stellar population (\cite{dg83}). Do these E+A galaxies have any
immediate connection to the present-day S0 population (Franx 1993)?

This paper is structured as follows. In \S \ref{data}, we outline the
selection and nature of the sample. In \S \ref{fpsec}, the fundamental
plane of the early-type galaxies is fit to the data, the S0s are
compared to the ellipticals, and the intrinsic scatter about the
fundamental plane is measured. In \S \ref{es0pops}, we discuss the
early-type galaxy $M/L_V$ ratios and their implications for the
stellar populations. In particular, we derive the evolution in $M/L_V$
with respect to the early-types in Coma; derive the extent to which
stellar populations are varying along the fundamental plane; and
determine the nature of the scatter in the fundamental plane and
color-magnitude relations. In \S \ref{spiral}, the fundamental plane
of the early-type spirals is discussed, and their $M/L_V$ ratios are
compared to those of the E/S0s. In \S \ref{models}, population
synthesis models are used to investigate the differences between the
stellar populations of the spirals and early-types, with the goal of
determining whether or not intermediate redshift spirals can evolve
into contemporary S0s. Our conclusions will be summarized in \S
\ref{conclusions}.


\section{Summary of Sample Selection, Observations, and Data
Reduction}
\label{data} 

We are currently studying the galaxy populations of three clusters in
detail, CL1358+62 ($z=0.33$), MS2053--04 ($z=0.58$), and MS1054--03
($z=0.83$), selected from the Einstein Medium Sensitivity Survey
(\cite{gioia}). During Cycle 5, a mosaic of twelve HST WFPC2 pointings
of CL1358+62 was taken in two filters, F606W and F814W. These data
were presented in van Dokkum \etal\ (1998a).

A large number of redshifts in the cluster has been compiled by Fisher
\etal\ (1998), who defined membership between $0.31461 < z < 0.34201$.
Using 232 cluster members within a $10'\times 11'$ field of view, they
found a velocity dispersion for the cluster of $1027^{+51}_{-45}$ \kms
(See Fisher \etal\ (1998) and Fabricant \etal\ (1999) for discussions
about the spatial and kinematic distributions broken down by
spectroscopic and morphological classifications). For fundamental
plane analysis, we randomly selected cluster members within the field
of view of the HST mosaic, to $R \le 21$ mag. The selection was
performed with an effort to efficiently construct multi-slit plates
for the Low-Resolution Imaging Spectrograph (\cite{okelris}). We used
three masks, with different position angles on the sky. The region of
maximum overlap is in the center of the cluster, and thus the FP
sample is concentrated towards the core of the cluster.

Morphology was not a factor in the selection of our sample. In the
random selection process, three E+A galaxies were included; these will
be compared with those cluster members that have normal, early-type
spectra. The sample is about $50\%$ complete for $R\le 20.5$ mag (see
\cite{fish} for details on the statistical completeness of the
original redshift catalog). Three galaxies fainter than $R=21$ mag
were added to test the quality of velocity dispersion measurements at
faint magnitude limits.

The spectroscopic reductions are detailed in Kelson \etal\ (1999a). In
total, we have central velocity dispersions for 55 galaxies in the
cluster. Kelson \etal\ measured central velocity dispersions within an
approximately $1'' \times 1''$ aperture and corrected the values for
aperture size to a nominal circular aperture of $D=3\Sec 4$ at the
distance of Coma. Kelson \etal\ (1999a) also showed that the kinematic
profiles of the CL1358+62 members are similar to nearby early-type
galaxies, ensuring that the aperture corrections derived from nearby
galaxies remain valid for the $z=0.33$ data. Furthermore, the E/S0s
and early-type spirals have similar kinematic profiles, over the
aperture that Kelson \etal\ (1999a) used, ensuring that the velocity
dispersion aperture corrections do not depend strongly on morphology.  

The derivation of structural parameters from the HST imaging is
discussed in Kelson \etal\ (1999b). Unfortunately, two of the galaxies
were imaged too close to the WFPC2 CCD edges to obtain reliable
structural parameters. Thus, there are a total of 53 cluster members
in our determination of the fundamental plane in CL1358+62.

As was mentioned earlier, three galaxies have E+A spectra as defined
by the criterion of Fisher \etal\ (1998), in which $\rm
(H\delta+H\gamma+H\beta)/3
> 4$ \AA\ and [OII] 3727 \AA\ $< 5$ \AA.
The E+A fraction of our sample (6\%) is representative of the cluster
(5\%; \cite{fish}). Two of the 53 galaxies show evidence of current
star formation activity, with Balmer and [OIII] emission (the cD,
\#375; and the Sb, \#234). Fisher \etal\ (1998) report an
emission-line galaxy fraction of $9\pm3\%$ in the field of the HST
mosaic down to $R= 21$ mag. In the sample presented here, the fraction
is about half that (a $2\sigma$ difference), though the emission-line
galaxies are more common in the outer parts of the cluster. Although
the cD shows evidence for emission in its central parts, we consider
it ``normal'' and spectroscopically early-type for the fundamental
plane.

Optical morphologies have been taken from Fabricant \etal\ (1999).
These authors classified several hundred galaxies in the HST mosaic
according to morphological type, $T$. The galaxies in our
high-resolution spectroscopic sample have $T\in \{-5,-4,-3,0,1,2,3\}$
(E, E/S0, S0, S0/a, Sa, Sab, Sb, respectively). The galaxy
morphologies are listed in Table \ref{table}, in which there are 11
Es, 6 E/S0s, 14 S0s, 13 S0/as, 6 Sas, 2 Sabs, and 1 Sb. This
distribution is quite similar to nearby massive clusters ({\it
e.g.\/}, \cite{oemler}). Note that the classification scheme employed
in Fabricant \etal\ (1999) is not identical to that applied in van
Dokkum \etal\ (1998a). The latter authors classified galaxies based on
whether or not they had any obvious disk structure, and did not
attempt to classify their galaxies according to the traditional Hubble
scheme.

The photometric parameters were derived from the deep WFPC2 data. The
surface photometry was transformed to Johnson $V$, redshifted to the
frame of the galaxies. Such a transformation is necessary to
facilitate a direct comparison of our sample with nearby galaxies.
Details of the transformation are given in van Dokkum \& Franx (1996)
and Kelson \etal\ (1999b). The formal uncertainties in the individual
structural parameters do not fairly represent the true errors in the
fundamental plane determination. Random and systematic errors in $r_e$
and $\langle \mu_e\rangle$ can be quite large, but their combined
error in the fundamental plane is quite small, at the level of a few
percent (see \cite{kelson99b} and, for example, the detailed analysis
presented in \cite{saglia}). The structural parameters were derived
using three different profiles: (1) the de Vaucouleurs $r^{1/4}$-law;
(2) generalized $r^{1/n}$-laws (\cite{sersic}); and (3) $r^{1/4}$-law
bulge plus exponential disk superpositions. Therefore, for each galaxy
we have three sets of structural parameters. In \S \ref{quant} and
\ref{residuals}, we investigate the utility of these different
profiles, for purposes of morphological classification, or for
determining accurate half-light radii and surface brightnesses.

The galaxy parameters used in the following analysis are listed in
Table \ref{table}. Included are the optical morphologies, the $n$
shape parameters, central velocity dispersions, $r_e$ and $\mu_e$ from
the $r^{1/4}$-law fit, $R$ magnitude from Fabricant \etal\ (1991), and
restframe $(B-V)$ colors from van Dokkum \etal\ (1998a). All three,
different sets of structural parameters are listed in Kelson \etal\
(1999b). They are not critical for the conclusions of this paper.


\section{The Fundamental Plane of Early-Type Galaxies}
\label{fpsec}

\subsection{Fitting the Fundamental Plane}

The fundamental plane is a power-law relation between effective
radius, $r_e$, central velocity dispersion, $\sigma$, and mean surface
brightness, $\langle I \rangle_e$, within $r_e$, of the following
form:
\begin{equation}
\log{r_e} = \alpha \log{\sigma} + \beta \log{\langle I\rangle_e } + \gamma.
\label{fpeq}
\end{equation}
For the present analysis, we follow the procedures of J\o{}rgensen
\etal\ (1996), who found the best-fit plane through the data by
minimizing the average absolute residual, perpendicular to the fitted
plane. This method has the advantage of being robust against outliers.
Similar to minimizing $\chi^2$ using a gradient search algorithm, we
search the parameter space of $\alpha$ and $\beta$ for the values
which minimize
\begin{eqnarray}
\chi &=& \sum {
|\log{r_e} - (\alpha \log{\sigma} + \beta \log{\langle I\rangle_e}
+\gamma)|
\over
\sqrt{1 +  \alpha^2+\beta^2} }\nl
&=& \sum { | \Delta \log{r_e}|
\over
\sqrt{1 +  \alpha^2+\beta^2} }
\end{eqnarray}
where the denominator arises from the rotation of the residuals in
$\log r_e$ to a vector normal to the plane. This choice is also
motivated, in part, by our desire to compare our results to the
extensive work by J\o{}rgensen \etal. In the fitting of a plane to
three dimensional data, with correlated and uncorrelated errors, no
single fitting procedure will accurately recover the underlying
relation. An accurate determination of the true physical relation
requires sophisticated modeling and knowledge of the underlying
distribution of galaxy parameters, all of which are beyond the scope
of this paper (\cite{efarmodel,models}). As was mentioned earlier,
J\o{}rgensen \etal\ (1996) found $\alpha = 1.16\pm 0.10$,
$\beta=-0.76\pm 0.04$ in Gunn $g$ using 109 galaxies in nearby
clusters. We adopt these values for the local slope of the fundamental
plane, as Gunn $g$ more closely matches the Johnson $V$-band, than
does Gunn $r$. Nevertheless, one should note that the slopes for the
local fundamental plane do not differ greatly between the two
bandpasses and our conclusions in this paper are not sensitive to this
choice.

Uncertainties in the values of $\alpha$, $\beta$, and $\gamma$ are
estimated by the bootstrap method ({\it e.g.\/}, \cite{beers}). Such
uncertainties represent the formal errors in the fit, when performed
in the manner we have adopted (minimizing the absolute residuals).
Systematic errors in the coefficients of the fundamental plane can
arise from many sources, such as the selection criteria, the
correlated measurement errors, and the choice of the fitting procedure
itself ({\it e.g.\/}, \cite{jfk96,models}). For example, differences
in fitting techniques can lead to small changes in the coefficients on
the order of $\pm 0.1$ or so. Thus, the fitting procedures of
J\o{}rgensen \etal\ (1996) were adopted partly so that this systematic
effect would be minimized in our measurement of evolution in the
coefficients. The systematic effects arising from selection biases and
correlated measurement errors are still being explored, and will be
published later (\cite{models}).

The scatter about the best-fit plane will be reported either in units
of $r_e$ at fixed $\sigma$ and $\langle I\rangle_e$, or in units of
$M/L_V$ at fixed $\sigma$ and $r_e$. Under the assumption of homology,
the observed quantities $r_e$, $\langle I\rangle_e$, and $\sigma$
relate to real physical properties of size, surface brightness, and
total second velocity moment, in the same manner for all early-type
galaxies. Thus, one can use the fundamental plane of the ellipticals
to transform the virial theorem
\begin{equation}
M/L\big|_{\rm obs}\propto \sigma^2 r^{-1} \langle I\rangle^{-1}_e
\end{equation}
into
\begin{equation}
M/L\big|_{\rm pred}\propto \sigma^{\alpha/\beta+2}
r_e^{-(1+\beta)/\beta}
\label{eq:mlpred}
\end{equation}
In Gunn $g$ the slopes for early-type galaxies of J\o{}rgensen \etal\
(1996) yield $M/L_g\big|_{\rm pred}\propto \sigma^{0.47\pm 0.15}
r_e^{0.32\pm 0.07}$

These equations enable us to express the fundamental plane residuals
in units of $M/L$ ratio:
\begin{eqnarray}
\Delta \log M/L &=& \log M/L\big|_{\rm obs} - \log
M/L\big|_{\rm pred}\cr
&=& (\log r_e-\alpha\log\sigma-\beta\log\langle I\rangle_e
-\gamma)/\beta\cr
&=& (\Delta \log r_e)/\beta
\label{eq:mlres}
\end{eqnarray}

Because the absolute residuals have been minimized, the \rms\ scatter
about any given fit is not. We therefore refrain from writing the
scatter as an \rms\ value, and instead report estimates of the
1-$\sigma$ scatter using $1.25\times $ the average absolute deviation,
either in $r_e$ or in $M/L_V$. For a Gaussian distribution of
residuals, this estimate for the scatter is equivalent to the standard
deviation. We find similar results when using the bi-weight statistic
(\cite{beers}), and conclude that our estimate of the scatter is
robust.

In constructing the figures, we express $r_e$ in units of kpc, and
$\langle I\rangle_e$, the mean surface brightness within an effective
radius, in units of $L_\odot/\rm pc^2$ in the $V$-band. For the
purposes of this paper, we use $H_0=65$ \kms\Mpc-1 and $q_0=0.1$,
defining a scale of 4.8 kpc/arcsec.

\subsection{The Fundamental Plane in CL1358+62}

\begin{figure}
\centerline{\vbox{
\hbox{
\mkfigbox{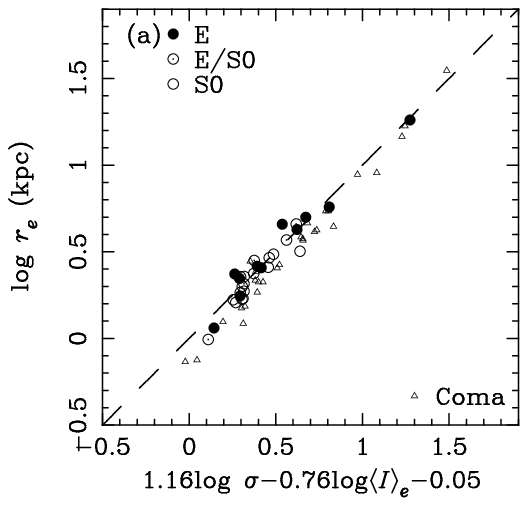}{0.4\textwidth}
\mkfigbox{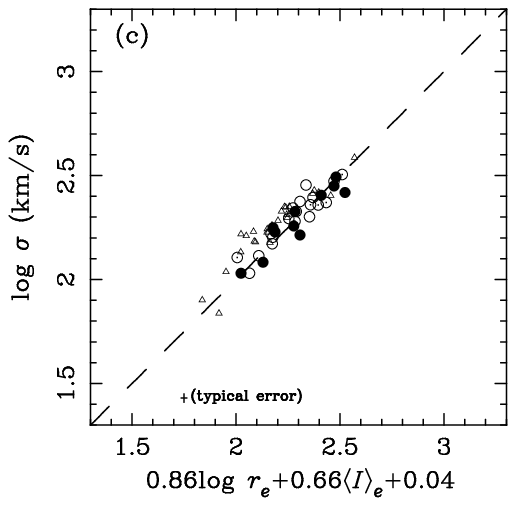}{0.4\textwidth}}
\hbox{
\mkfigbox{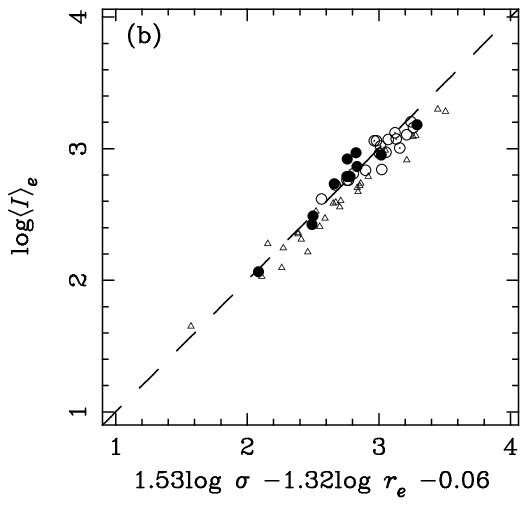}{0.4\textwidth}
\mkfigbox{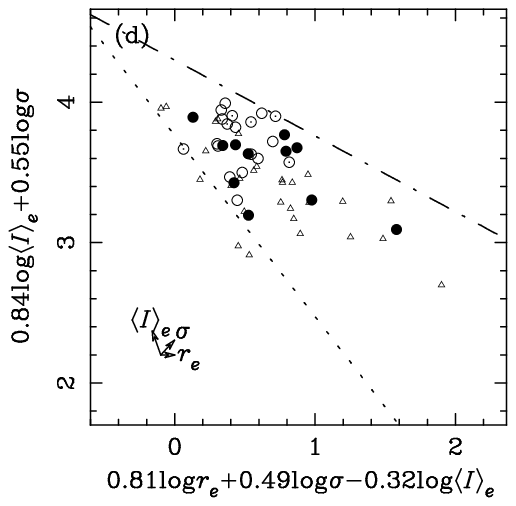}{0.4\textwidth}}}}
\figcaption[fig1a.ps, fig1b.ps, fig1c.ps, fig1d.ps]{
The fundamental plane of early-type galaxies in CL1358+62. In (a-c) we
show the long, intermediate, and short edge-on views and in (d) the
face-on distribution. The early-type galaxies clearly form a tight
relation in (a-c). The dashed lines show the fit to the elliptical
galaxies of CL1358+62. In (c), we show the size of a typical error.
The absence of galaxies in the lower left region of (d) is attributed
to the magnitude limit of the sample, denoted by the dotted line. The
absence of galaxies in the upper right is referred to as the ``Zone of
Avoidance'' (\cite{bbf93}). The CL1358+62 galaxies are shown by the
large symbols indicated in (a). The Coma $V$-band sample is shown by
small triangles (\cite{jfk96}).
\label{view1}}
\end{figure}

In Figure \ref{view1}, we show the fundamental plane of early-type
galaxies. The long, intermediate, short edge-on projections, and the
face-on projection are shown in Fig.~\ref{view1}(a-d). There is a
tight relation, offset from the relation defined by the early-type
galaxies of Coma (\cite{jfk96}). In \ref{view1}(d), the ``Zone of
Avoidance'' (\cite{bbf93}) is the region in the upper right which
contains no galaxies. Note that the CL1358+62 early-type galaxies have
a different distribution than the Coma galaxies: the CL1358+62
galaxies lie at slightly higher surface brightnesses. We will come
back to this point later.

The fundamental plane of the 30 E, E/S0, and S0 galaxies (not
including any galaxies with E+A spectra) is described by
\begin{equation}
r_e \propto \sigma^{1.31\pm 0.13} \langle I\rangle_e^{-0.86\pm 0.10}
\label{fundplan}
\end{equation}
The 1-$\sigma$ scatter is only $\pm 14\%$
in $r_e$. Using Equation \ref{eq:mlres}, one obtains
\begin{equation}
M/L_V\big|_{\rm pred}
\propto \sigma^{0.48\pm 0.18} r_e^{0.16\pm 0.14}
\label{mlpred}
\end{equation}
with a scatter of about $\pm 16\%$ in $M/L_V$ ratio. The scatter in
$M/L_V$ is $18\%$ when using the locally defined Gunn $g$ slopes.

The values obtained above for the slope do not significantly differ
from that found locally, either in Gunn $g$ or Gunn $r$ (see above).
Comparing Equation \ref{mlpred} to the local expectation in Gunn $g$,
one finds that the slope of the fundamental plane has not
significantly evolved from the present-day:
\begin{eqnarray}
\log M/L\big|_{z=0.33} -
\log M/L\big|_{z=0} \propto \qquad\qquad \qquad\cr
 (0.01\pm 0.23)\log \sigma - (0.16\pm 0.16)\log r_e
\end{eqnarray}

Both van Dokkum \& Franx (1996) and Kelson \etal\ (1997) noted that
the intermediate redshift FP slope may not be the same as that found
locally. The advantage of the work presented here is the depth and
size of the sample, which allows for a more detailed breakdown by
morphological type. Kelson (1998, 1999) show that a shallow magnitude
cut-off in the selection process can lead to a flattening of the
fitted fundamental plane slope, similar to other Malmquist-like
regression biases. Such an effect probably led to the mild change in
slope observed by Kelson \etal\ (1997) and van Dokkum \& Franx (1996).


\subsection{Quantitative Morphologies}
\label{quant}

We tested the robustness of the slope to the morphological
classifications. Using the Sersic (1968) profile shape parameters $n$
as a classification tool, the plane of the 30 galaxies with $n\ge 4$
are best fit by $\alpha=1.19\pm 0.23$ and $\beta=-0.76\pm 0.12$. Using
the 1D bulge-disk decompositions, the plane of the 25 galaxies with
bulge fractions greater than 90\% are best fit by $\alpha=1.19\pm
0.17$ and $\beta=-0.76\pm 0.06$. Recall that the visually classified
E/S0s galaxies produced slopes of $\alpha=1.31\pm 0.13$ and
$\beta=-0.86\pm 0.10$. We conclude that our measured fundamental plane
slopes are not affected by uncertainties in the morphological
classifications.

\bigskip

Since the slopes we have determined for the CL1358+62 FP do not differ
significantly from that found by J\o{}rgensen \etal\ (1996), as given
earlier, we continue our analysis using the locally defined values of
$\alpha=1.16$ and $\beta=0.76$. The remaining conclusions of this paper
are not sensitive to the adoption of these values over the ones
measured above.


\subsection{Elliptical Galaxies versus Lenticulars}

For nearby S0s, the zero-point and scatter of the fundamental plane is
the same as that for ellipticals (\cite{jfk96}). Using either the
fundamental plane coefficients defined by the 30 CL1358+62
early-types, or the ones defined locally, the 11 ellipticals have the
same zero-point as the 13 non-E+A S0s. This is fully consistent with
what has been found in nearby clusters.

There remains, however, the question of whether the slope of the
fundamental plane of the S0s is the same as that of the ellipticals.
The 11 visually classified ellipticals yield $\alpha=1.25\pm0.26$ and
$\beta=-0.71\pm0.16$ with a 1-$\sigma$ scatter of $\pm 10\%$ in $r_e$.
There appears to be no significant change in the slope by removing the
one elliptical fainter than the magnitude limit and exclusion of the
cD does not significantly affect the slope either. By fitting a plane
to the 13 non-E+A S0s in CL1358+62, we find $\alpha=1.12\pm 0.36$,
$\beta=-0.86\pm 0.22$, with a scatter of $\pm 8\%$. The difference in
scatter is not statistically significant. Because the distinction
between E and S0 can become more difficult at higher redshift, we have
also divided the sample of 53 galaxies by bulge-to-disk ratio, derived
from fitting two-component $r^{1/4}$-law bulge and disk growth curves
to surface photometry (\cite{kelson99b}). No significant difference in
the fundamental plane slopes are seen for any of these subsets.

As a result of these tests, we conclude that differences between the
coefficients of the FP relations of the S0s and Es, if any, are too
subtle to be accurately measured with the current sample.


\subsection{The Intrinsic Scatter of the Early-Type Galaxies}
\label{obsscat}

In cosmological models involving hierarchical merging of gas-rich
systems, early-type galaxies are expected to have some intrinsic
scatter in the properties of their stellar populations
(\cite{kauf95},1996). Such variations would manifest themselves in a
scatter in the colors, absorption line strengths, and $M/L_V$ ratios
of early-type galaxies at a given luminosity, galaxy mass, or velocity
dispersion (\cite{sandagecm,terlevich,faber87}).

Evidence for a spread in the properties of stellar populations has
been measured using the scatter about the color-magnitude relation
({\it e.g.\/}, \cite{bower,ellis97,stanford}). We have measured the
scatter in colors about the color-magnitude relation for the galaxies
in our sample and find an {\it intrinsic\/} 1-$\sigma $ scatter in
color, at a given mass, of $\pm 0.018$ mag (see \cite{vdcm}). Using
the GISSEL96 (\cite{mod}) single burst stellar populations with a
Salpeter (1955) initial mass function (IMF), and assuming that
variations in age are the cause of these residuals, this scatter in
color is equivalent to a scatter in $M/L_V$ of $\pm 13\%$.

Equation \ref{eq:mlres} showed that variations in $M/L_V$ ratios of
stellar populations, at a given $r_e$ and $\sigma$ will lead to
scatter in the fundamental plane. Residuals from the fundamental plane
can be caused by a number of other effects as well, such as variations
in elliptical galaxy shapes, metal abundance, and dust content. In
contrast, the scatter in the color-magnitude relation may only be due
to variations in stellar populations and/or dust content.  Some have
also suggested that correlations between age, other stellar population
properties, and/or galaxy structural parameters may be conspiring to
reduce the scatter in the fundamental plane and color-magnitude
relations (cf. \cite{wtf,trager}). J\o{}rgensen \etal\ (1996) have
limited to what extent random projection can inflate the scatter of
the fundamental plane to less than 6\% in $M/L$ ratio. Together, the
scatter in the fundamental plane and color-magnitude relations can be
used to constrain the properties of the stellar populations.

The ellipticals and S0s have an observed 1-$\sigma$ scatter about the
FP of $\pm 16\%$ in $M/L_V$. The observational errors in our data are
a small component of the observed scatter. As shown in Kelson \etal\
(1999a), the formal velocity dispersion uncertainties range from $\pm
2\hbox{-}7\%$. Because of template mismatch, the true errors are more
like $\pm 4\hbox{-}8\%$. The fundamental plane parameters, $r_e\langle
I\rangle_e^{-0.8}$, have random errors of $\pm 3\%$
(\cite{kelson99b}). Thus, the contribution to the observed scatter
from measurement errors is approximately $\pm 7\%$ in $r_e$ ($\pm 8\%$
in $M/L_V$). Removing this in quadrature gives an intrinsic scatter in
$M/L_V$ of $\pm 14\%$. That the intrinsic scatter in $M/L_V$ should
agree so well with the naive expectation from the scatter in colors is
remarkable. Moreover, the intrinsic scatter of $14\%$ is an upper
limit to the scatter in the stellar populations because there may be
other sources of scatter in the observed $M/L_V$ ratios, due to
variations, for example, in galactic structure.


\section{Implications for the Stellar Populations of Early-Type
Galaxies}
\label{es0pops}

\subsection{The Evolution of the Zero-point for the Fundamental Plane of
Early-Type Galaxies}
\label{evolution}

The zero-point of the fundamental plane is related to the mean $M/L$
ratio of a sample of early-type galaxies. Evolution in the
luminosity-weighted mean properties of stellar populations therefore
implies evolution in the FP zero-point. Although modest evolution of
the FP zero-point has already been demonstrated
(\cite{vdf96,kelson,vdfp98}), these new data can improve upon the
accuracy of the results in Kelson \etal, and test the sensitivity to
small number statistics and sample biases.

We use a sample of early-type galaxies in Coma to define the
present-day zero-point of the fundamental plane. These data are a
subset of the large Gunn $r$ sample in J\o{}rgensen \etal\ (1996) that
also had the required color information to transform the surface
brightnesses to the $V$-band. This $V$-band Coma sample is shown in
Figures \ref{view1}(a-d) by the small triangles. Note that the
CL1358+62 early-types are shifted by a small amount with respect to
the Coma early-types. The trend appears to be towards increasing
surface brightnesses, suggesting that at least some of the shift in
$M/L$ derives from luminosity evolution.

If the ranges of galaxy masses in Coma and CL1358+62 are substantially
different, which was the case for the small samples in van Dokkum \&
Franx (1996), Kelson \etal\ (1997) and van Dokkum \etal\ (1998b), then
the offset in $M/L_V$ can be very sensitive to the adopted slopes of
the fundamental plane. This sensitivity arises because the fundamental
plane is essentially a relation between $M/L_V$ ratio and galaxy mass.
If one compares two samples of galaxies with substantially different
mass functions, then any derived offset in $M/L_V$ is going to be a
combination of the true offset, the differences in FP slope, and the
difference in the mean mass of the sample. By minimizing the
differences between the mass distributions, one eliminates this latter
term, and the results become insensitive to the slope of the FP.
Therefore, when deriving the offsets, we restrict the local
calibrating sample to a similar mass range as in CL1358+62. Such a
restriction on the local sample implies that any measured evolution is
relevant for galaxies within a specific range of masses.

\begin{figure}
\centerline{ \mkfigbox{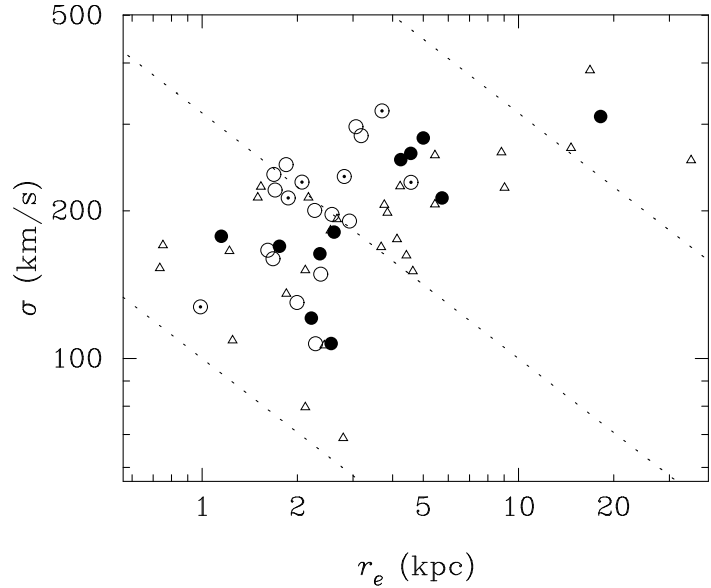}{0.5\textwidth}}
\figcaption[fig2.ps]{
The $\log\sigma$-$\log r_e$ plane in CL1358+62 vs. the Coma sample
(\cite{jfk96}). The symbols are as in Figures \ref{view1}. Contours of
constant mass ($5\sigma^2r_e/G=10^{10,11,12}M_\odot$, \cite{bbf92})
are shown by the dotted lines. The distribution of early-type galaxy
masses at $z=0.33$ appears remarkably similar to the $V$-band
early-type sample in Coma.
\label{mass}}
\end{figure}

In Figure \ref{mass}, we show the distributions of the CL1358+62
galaxies overlayed with the Coma $V$-band sample in the
$\log\sigma$-$\log r_e$ plane. The dashed lines show contours of
constant mass ($5G^{-1}\sigma^2 r_e=10^{10,11,12}M_\odot$,
\cite{bbf92}). The distributions are remarkably similar, ensuring that
our measurement of the $M/L_V$ evolution will not be biased by any
differences between the two distributions of galaxy masses.

Using the 30 early-type galaxies, we find that the $M/L_V$ ratios, for
a given $r_e$ and $\sigma$, are offset by $\Delta \log
M/L_V=-0.13\pm0.03$ ($q_0=0.1$; $-0.10\pm 0.03$ dex for $q_0=0.5$).
The errors listed are the formal errors in the offset. We test the
sensitivity of this result to the morphological classifications by
using only the Es and find no significant difference. Changing the
adopted slope from the J\o{}rgensen \etal\ (1996) Gunn $g$ values to
those for Gunn $r$ changes the results by $-0.02$ dex. Using the
values found in in \S \ref{fpsec} also produces no significant change
in the result as well. Kelson \etal\ (1997) had found $\Delta \log
M/L_V = -0.17 \pm 0.05$ dex. The difference can be attributed to
improvements in the size and depth of the sample.

We can use the simple, single-burst models of \S \ref{models} to infer
the relative age difference between the Coma and CL1358+62
early-types.  Assuming that the early-type galaxies in both clusters
have had similar formation histories and adopting a Salpeter IMF,
the CL1358+62 early-type galaxies have ages in the mean which are
$70\pm 5\%$ ($q_0=0.1$; $76\pm 5\%$ for $q_0=0.5$) of the mean ages
of their counterparts in Coma, for galaxies of the same $r_e$ and
$\sigma$. These uncertainties are the internal errors alone. The
external errors account for another 5-10\%. Thus, \about $1/4$ of the
lifetimes of cluster ellipticals has passed during the last 30\% of
the age of the universe.


\subsection{The Variation in $M/L_V$ along the Fundamental Plane}
\label{fpvary}

In the previous section we showed that the intrinsic scatter in the
fundamental plane is very low, at \about 14\% in $M/L_V$ ratio. The
implication is that the $M/L_V$ ratios of E/S0s are very tightly
correlated with their structural parameters in Cl1358+62. In this
section, we introduce simple models for stellar populations to address
the implications of such a correlation.

Tinsley \& Gunn (1976) first showed that the $M/L_V$ ratio of a
single-burst stellar population evolves as
\begin{equation}
M/L_V \propto t^{1.2-0.28x},
\label{tinsley}
\end{equation}
where $t$ is the time since the formation epoch, and $x$ is the slope
of the initial mass function. They found very little dependence
on metallicity, $Z$. This model predicts $\log M/L_V \propto 0.82 \log
t$ for an IMF with $x=1.35$ (\cite{imf}). For composite stellar
populations, $t$ and $Z$ are luminosity-weighted mean ages and metal
abundances.

For our analysis, we turn to the GISSEL96 (\cite{mod}) stellar
population synthesis models. We choose to use bivariate least-squares
fits to the model curves in order to use simple analytical expressions
in the analysis. The models of single-burst stellar populations older
than $t \ge 2$ Gyr, with a Salpeter (1955) IMF from $m=0.1$-$125
M_\odot$ are well fit by
\begin{equation}
\log M/L_V = -6.74 + 0.84 \log t + 0.56 \log Z + 0.076 (\log Z)^2,
\label{bzml}
\end{equation}
and
\begin{equation}
(B-V) = -1.19 + 0.27 \log t + 0.34 \log Z + 0.027 (\log Z)^2,
\label{bzbv}
\end{equation}
where $t$ is in Gyr, and $Z$ is the metal abundance (solar abundance
ratios), between $Z=4\times 10^{-4}$ to $5\times 10^{-2}$. For
composite stellar populations, the properties of age and metallicity
are mean properties of the luminosity-weighted sum of the stellar
populations within each galaxy.

\begin{figure}
\centerline{\mkfigbox{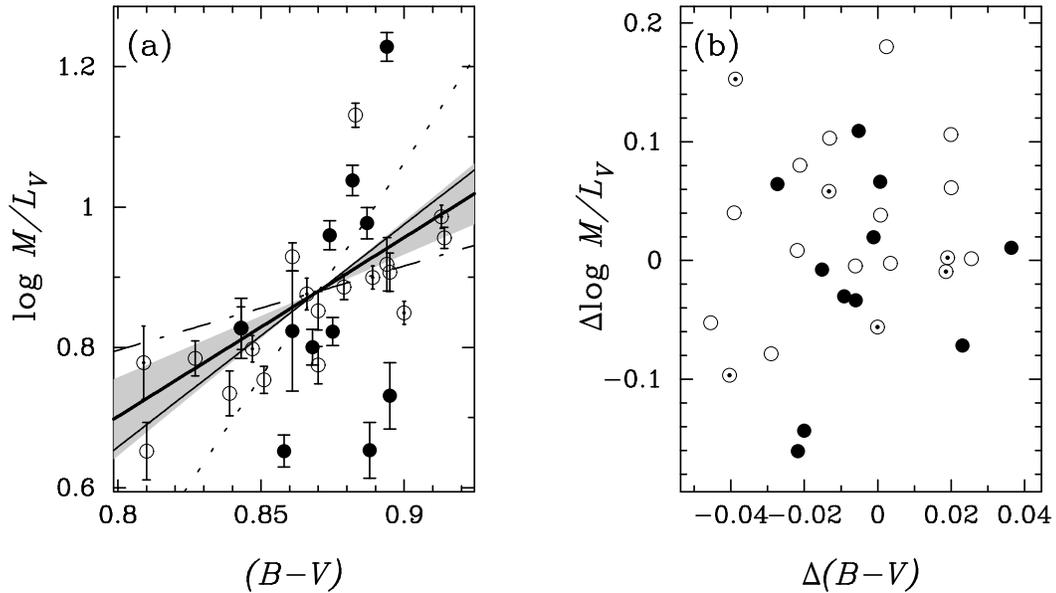}{0.75\textwidth}}
\figcaption[fig3.ps]{
(a) $M/L_V$ ratios are shown plotted against $(B-V)$ colors for the
early-type galaxies. A Spearman rank test indicates a $99.9\%$
probability that $M/L_V$ ratios are correlated with galaxy color. The
linear-least squares fit gives a slope of $2.55\pm 0.79$, as shown by
the thick solid line. The 1-$\sigma$ uncertainties are shown by the
enclosed gray region. The thin solid line is the slope expected for
systematic age variations, the thin dash-dot line for pure metallicity
variations. The slope predicted by the Worthey (1994) "3/2" rule is
shown by the thin dotted line. Typical color uncertainties are $\pm
0.011$ mag (\cite{vdcm}). In (b) fundamental plane residuals are
plotted against the residuals from the $(B-V)$ color-magnitude
relation and no significant correlation is seen.
\label{dbv_es0}}
\end{figure}

The implication of Equations \ref{bzml} and \ref{bzbv} is that $M/L_V$
ratios should be correlated with galaxy color. This correlation is
explicitly shown in Figure \ref{dbv_es0}(a). Because the underlying
correlations of $M/L_V$ and $(B-V)$ with $r_e$ and $\sigma$ have not
been removed, the figure is a combined projection of the fundamental
plane and color-magnitude relations. A Spearman rank correlation test
indicates a probability of 99.9\% that the $(B-V)$ colors and $M/L_V$
ratios of the E/S0s are correlated.

A linear least-squares fit to the points in the figure gives $\log
M/L_V \propto (2.55\pm 0.79)\times (B-V)$, shown by the thick solid
line. The shaded region indicates the $\pm 1$-$\sigma$ errors in the
slope. The uncertainty in the slope is large because the scatter in
the diagram is large, and because the errors for individual
measurements of $M/L_V$ are also large.

As stated above, the time-evolution of $M/L_V$ is governed largely by
the shape of the IMF (\cite{tg76}). For stellar populations of
constant metallicity Equations \ref{bzml} and \ref{bzbv} imply a
relation between $M/L_V$ ratio and color of $\log M/L_V \propto
3.1\times (B-V)$. Thus, if the correlation in Figure \ref{dbv_es0}(a)
is the result of a systematic variation in galaxy age, then one would
expect a slope indicated by the thin solid line. Under this
assumption, any departure of the data in Figure \ref{dbv_es0}(a) from
a slope of 3.1 implies an error in the adopted IMF. Using the IMF
dependence of Equation \ref{tinsley}, we make the following simple
modification to Eq. \ref{bzml}:
\begin{eqnarray}
\log M/L_V=
-6.74+(1.22-0.28x)\log t\qquad\cr
+0.56 \log Z + 0.076 (\log Z)^2.
\label{bzmod}
\end{eqnarray}
Equation \ref{bzbv} is insensitive to the shape of the IMF because the
dependence of both the $M/L_V$ and $M/L_B$ ratios on $x$ is nearly
identical (\cite{tg76}). Put another way, $(B-V)$ color is related to
the temperature of the main-sequence turn-off stars, which itself is
independent of the shape of the IMF.

Using the least-squares fit to Figure \ref{dbv_es0}(a), the slope of
the IMF is constrained to be $x=1.9\pm 0.8$. The implication of this
IMF is that $M/L_V$ evolves according to $\log M/L_V \propto (0.67\pm
0.22)\log t$. Given that $\log M/L_V \propto 0.24\log M + 0.08\log
r_e$, one concludes $\log t\propto {\sim 0.35_{-0.08}^{+0.18}}\log M$.
The implication is that the mean luminosity-weighted age of elliptical
galaxies in CL1358+62 with $\sigma=100$ \kms\ is $45_{-15}^{+8}\%$ of
the mean luminosity-weighted age of the $\sigma=300$ \kms\
ellipticals. For a Salpeter (1955) IMF, this ratio of ages is 50\%.

If a systematic trend in metallicity is causing the correlation in
Figure \ref{dbv_es0}(a), then, near solar metallicities, the slope
should be $\log M/L_V \sim 1.2 (B-V)$, shown by the thin dash-dot
line. The fitted slope is discrepant from that expected from
metallicity effects by less than 2-$\sigma$. Worthey (1994) and others
have suggested that luminosity-weighted ages and metallicities are
correlated such that $\Delta \log t/\Delta \log Z \propto -3/2$. The
expected curve is based on this relation is shown as the thin dotted
line in Figure \ref{dbv_es0}(a). No particularly strong constraints
can be made with the current data if both age and metallicity are
varying.

The correlation of $M/L_V$ ratio with $(B-V)$ color can only be
related to the stellar populations. Do these systematic variations
in the stellar populations fully account for the observed slope of the
fundamental plane? By correcting the surface brightnesses for the
fitted correlation of $M/L_V$ with $(B-V)$ color, and fitting a new
plane to the corrected data, we find:
\begin{equation}
\log r_e \propto (1.84 \pm 0.27) \log \sigma - (0.89\pm 0.17)
[\log \langle I\rangle _e + 2.55 (B-V)]
\label{eq:fpcor}
\end{equation}
This fit implies that the early-type galaxies in CL1358+62 are
consistent with notion that they are a family of homologous objects.
The uncertainties are large, in part, because we have assumed that the
correlation between $M/L_V$ ratio and $(B-V)$ color applies for the
entire sample of early-type galaxies in the cluster. Individual
galaxies may have histories which deviate substantially from the mean
star-formation history of E/S0s, and thus be poorly modeled by the
fitted correlation. For example, we cannot assume that all early-type
galaxies have constant metallicity at fixed age, or vice versa. There
is some naturally occurring dispersion in the mean properties of the
stellar populations of E/S0s and this dispersion is reflected as
scatter in Figure \ref{dbv_es0}(a).

More data, such as from absorption-line strengths and larger sample
sizes, are required in order to reduce the uncertainties. Furthermore,
more careful modeling of the observations will be required in order to
fully account for the fitting biases and measurement errors
(\cite{thesis}, 1999).


\subsection{Understanding the Scatter in the FP and CM Relations}
\label{sourcescat}

Residuals in the fundamental plane and color-magnitude relations are
expected to be related through Equations \ref{bzbv} and \ref{bzmod}.
When the residuals in $\log M/L_V$ and $B-V$ are plotted against each
other, as in Figure \ref{dbv_es0}(b), no obvious correlation is seen
for the E/S0s. Residuals from Eq. \ref{eq:fpcor} do not show any
significant correlation with the color-magnitude residuals either.

Perhaps measurement errors, or other sources of scatter besides the
stellar populations are large enough to obscure any intrinsic
correlation between the residuals from the relations? One source of
uncertainty that has been ignored is the intrinsic correlation of
$r_e$ and $\langle I\rangle_e$, and how it propagates into errors in
$M/L_V$ ratio. The $M/L$ ratios are expected to scale as $(r_e\langle
I\rangle_e)^{-1}$ for a homologous family of objects, yet the error
correlation in the photometric structural parameters indicates a
correlation of $r_e\langle I\rangle_e^{\about 0.75}$ (Kelson \etal\
1999b, Kelson \etal\ 1999c). Given the typical random uncertainties in
half-light radii of approximately $\delta \log r_e\approx 0.15$ dex,
the corresponding, random uncertainties in $M/L_V$ are $\delta \log
M/L_V \approx 0.05$ dex. This additional source of error does not have
this large an impact on the fundamental plane scatter, however,
because the underlying correlation between $M/L_V$ and $r_e$ is not
orthogonal to the error correlation between $r_e$ and $\langle
I\rangle_e$. The error correlation is probably within $\pm 30^\circ$
of the underlying correlation between $M/L_V$ and $r_e$
(\cite{models}). Therefore the $\pm 0.05$ dex random scatter in the
$M/L_V$ ratios is diminished in its contribution to the fundamental
plane scatter. The effect of this error correlation is to mix up the
distribution of galaxies within the fundamental plane, leaving
correlations between $\Delta \log M/L_V$ and $\Delta (B-V)$ difficult
to observe.

Because of the observational effects discussed in the previous
paragraph, one's ability to interpret the scatter of the FP and CM
relations is limited. While we cannot directly observe correlations
between the FP and CM residuals, we can use the intrinsic scatter in
the FP and CM relations to place constraints on the stellar
populations. Several assumptions are made in the following analysis.
In particular, we assume that the intrinsic scatter in both relations
is due to a dispersion in the luminosity-weighted mean properties of
the stellar populations at fixed $r_e$ and $\sigma$.

The {\it intrinsic\/} scatters of the FP and CM relations were found
to be $\sigma_{\log M/L_V} = \pm 0.060\pm 0.012$ dex, and
$\sigma_{(B-V)} = \pm 0.018 \pm 0.005$ mag. For the scatter in
$M/L_V$, the observed value of is an upper limit for the stellar
populations because the scatter we measure in $M/L_V$ may in part be
due to other factors, such as might be caused by random variations in
galaxy shape.

We now assume that the scatter in colors is due solely to variations
in stellar population ages, at fixed mass, and investigate the
ramifications. The scatter in colors implies a 13\% scatter in
luminosity-weighted ages. Using the analytical models of the previous
section, one expects
\begin{equation}
\Delta \log M/L_V \propto (4.52\pm 1.0x) \times \Delta (B-V)
\label{eq:age}
\end{equation}
Using $x=1.35$, the scatter in colors is equivalent to a 13\% scatter
in the $M/L_V$ ratios.

The scatter in the two relations is remarkably consistent with a
scatter in luminosity-weighted ages, at fixed mass. Using simple
single-burst models of stellar populations, and assuming a Gaussian
distribution of residuals in $\log M/L_V$, we can draw some
conclusions from the scatter in $M/L_V$ and $(B-V)$. If the mean epoch
of star-formation for cluster ellipticals is $z=2$, then the
1-$\sigma$ scatter of 15\% in the luminosity-weighted ages implies
$\pm 1$-$\sigma$ interval for the epochs of star-formation of $1.4
\simlt z\simlt 3.5$. The $\pm 2$-$\sigma$ limits are $1 \simlt z\simlt
9$. Thus, 98\% of the star-formation in ellipticals was finished by
$z\approx 1$. These conclusions were based on an IMF of $x=1.35$, a
cosmology with $q_0=0.1$, and a mean formation redshift of $z=2$ that
is independent of velocity dispersion.

By measuring the evolution of the scatter to higher redshifts, one can
place stronger constraints on the star-formation rates and histories
of cluster ellipticals and S0s. If there exist systematic variations
in the mean luminosity-weighted ages of early-type galaxies along the
fundamental plane, {\it e.g.\/}, \S \ref{fpvary}), then one requires
more detailed modeling to derive the star-formation histories
(\cite{models}).


\section{Extending the Fundamental Plane to Early-Type Spirals}
\label{spiral}

In \S \ref{data}, we stressed that morphological information was
explicitly disregarded in selecting the spectroscopic sample discussed
in this paper. In this section, we use the fundamental plane as a tool
to place the later-type galaxies in the context of the evolution of
cluster galaxies in general. By doing so, the cluster populations can
be studied without biasing the results exclusively to the oldest
elliptical and lenticular galaxies. This aspect of our program makes
it unique among studies of high-redshift clusters ({\it e.g.\/},
\cite{ellis97,ziegler,pahrethesis,pahre98}). We use the early-types in
the comparison, as nearby fundamental plane data on spirals has not
been collected in as systematic and homogeneous manner as has been for
early-type samples.

\begin{figure}
\centerline{\vbox{
\hbox{
\mkfigbox{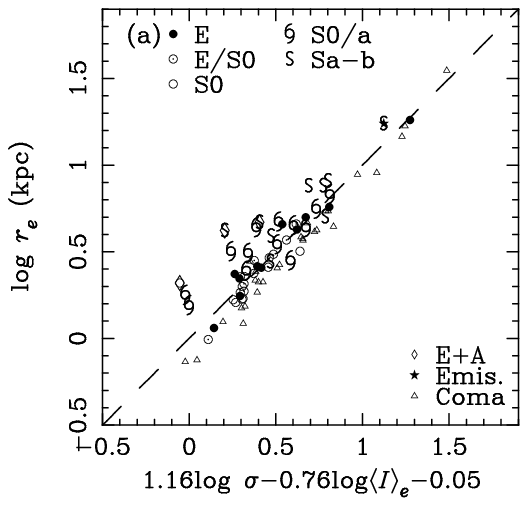}{0.4\textwidth}
\mkfigbox{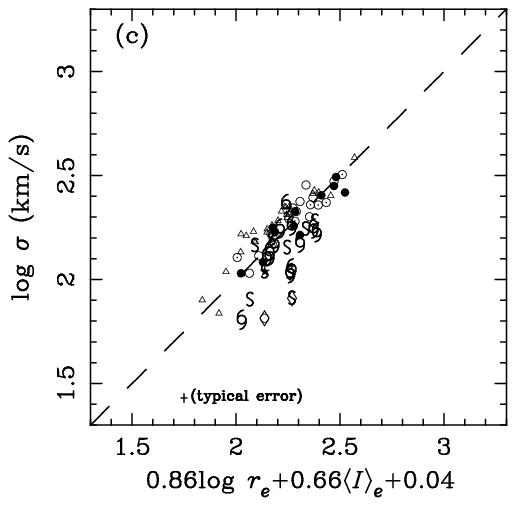}{0.4\textwidth}}
\hbox{
\mkfigbox{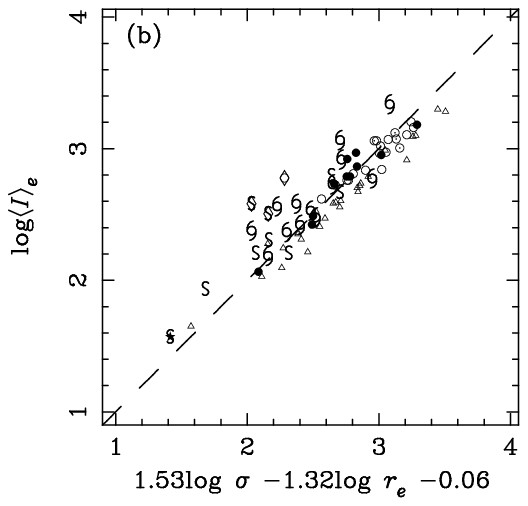}{0.4\textwidth}
\mkfigbox{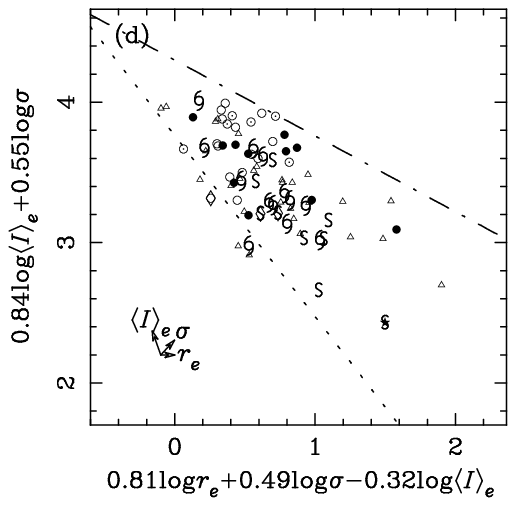}{0.4\textwidth}}}}
\figcaption[fig4a.ps, fig4b.ps, fig4c.ps, fig4d.ps]{
(a-d) The full sample of 53 galaxies are shown together. Note that the
later-type galaxies are systematically brighter than the plane of the
early-types.
\label{view2}}
\end{figure}

In Figure \ref{view2}, we expand the fundamental plane of early-types
to include all 53 galaxies in the sample. Recall that the sample of 53
galaxies contains 22 galaxies of morphological type S0/a and later. It
is clear from the figure that these early-type spirals show large
scatter compared to the E/S0s (shown by the smaller symbols). Using
all 53 galaxies, one obtains a plane with $\alpha=1.10\pm 0.20$ and
$\beta=-0.83\pm 0.08$. These values of the slope are not significantly
different than what was found earlier. Assuming that all of the
galaxies are homologous, the 1-$\sigma$ scatter is equivalent to $\pm
28\%$ in $M/L_V$. The scatter in the full sample is nearly twice as
large as that of the early-types alone.

There are 22 galaxies classified as S0/a and later. These galaxies
follow a fundamental plane with $\alpha=0.66\pm 0.29$,
$\beta=-0.64\pm0.06$, with a scatter of about $\pm 32\%$ in $M/L_V$.
This plane is a significant departure from that of the ellipticals.
Excluding the E+As or the emission-line galaxy \#234 does not affect
the slopes or scatter significantly. This large scatter implies that
the early-type spirals are very inhomogeneous compared to the E and S0
galaxies. By inhomogeneous, we refer to a dispersion in their observed
$M/L_V$ ratios at a given location along the FP. We can conclude that
the large scatter seen in the full sample of galaxies is partly due to
the inhomogeneity in the properties of the early-type spirals, and
partly due to the fact that those galaxies follow a different plane
than that of the E and S0 galaxies.

The difference in slope for the spirals, compared to the early-types,
may have several sources: (1) additional rotational support, (2)
systematic errors in their structural parameters, or (3) variations in
their stellar populations. Any or all of these may be in effect, but
they must be correlated with the structural parameters in order for
the fundamental plane of the spirals to be so well-defined.
Inhomogeneities in any of these properties are likely to be be present
as well, given the large scatter shown above.

While we do expect $M/L_V$ ratios, and thus the fundamental plane, to
differ between elliptical and spiral galaxies, the assumption of
homology may not be valid for our entire sample. For example, many of
the galaxies show strong rotation (\cite{kelson99a}), or large
deviations from an $r^{1/4}$-law surface brightness profile.


\subsection{The Spirals in the Context of the Early-Types}
\label{residuals}

We have shown that spirals follow a fundamental plane which is
systematically different from ellipticals and S0s. We now proceed to
investigate whether this change in slope is due to structural
(``non-homology'') or stellar population effects. In Figure
\ref{deviate1} we show the residuals from the fundamental plane of the
early-types plotted against (a) morphological type, (b) bulge fraction
of total light, $BF$, (c) the Sersic (1968) $n$ shape parameter, (d)
apparent ellipticity, $\epsilon$, (e) velocity dispersion, and (f)
residual from the color-magnitude relation, $\Delta (B-V)$
(\cite{vdcm}). The figures show a clear trend of fundamental plane
residual with galaxy morphology velocity dispersion, and with galaxy
color as well.

There are two possibilities for the correlations seen in Figure
\ref{deviate1}. Perhaps there are inherent structural differences
between the classes of E/S0s and early-type spiral galaxies, or
perhaps the $M/L_V$ ratios of the spirals deviate from the fundamental
plane of the early-types because of stellar population differences.
We now systematically determine which of these options is relevant for
our data.


\subsubsection{Bulge-to-Disk Ratio}

The figure shows a correlation with morphology, using $T$, $BF$, and
$n$. Early-type spirals clearly show large scatter, and systematically
negative residuals. Galaxies with low bulge fraction of total light
are also systematically offset from the galaxies with large bulge
fractions. Using the Sersic (1968) profile shape parameter $n$, one
finds that galaxies with low values of $n$ have, on average, lower
$M/L_V$ ratios, for a given $r_e$ and $\sigma$, and larger scatter in
those $M/L_V$ ratios, compared to galaxies with large values of $n$.

We tested whether using a pure $r^{1/4}$-law surface brightness
profile to derive structural parameters might have caused the
additional scatter and tilt for the FP of the spirals. Kelson \etal\
(1999b) derived several sets of structural parameters using
$r^{1/n}$-law profiles and superpositions of an $r^{1/4}$-law bulge
plus exponential disk. {\it Their results show that the combination of
$r_e$ and $\langle I\rangle_e$ that enters the fundamental plane was
very stable, irrespective of the choice of profile shape.}

By using the structural parameters derived from the $r^{1/n}$-law
profiles, or bulge-plus-disk superpositions, we find no significant
difference in the fundamental plane from that obtained earlier using
the de Vaucouleurs structural parameters. Given the stability of the
combination of $r_e$ and $\langle I\rangle_e$ that appears in the
fundamental plane, as shown in Kelson \etal\ (1999b), we conclude that
errors in the structural parameters are unlikely to have resulted in
any inhomogeneities in the observed properties of the CL1358+62
sample. Therefore for the remainder of the paper, we exclusively use
the de Vaucouleurs profile structural parameters.


\subsubsection{Projection}

If differences in galactic structure are causing large, systematic
residuals in $M/L_V$ ratio with respect to the fundamental plane of
the ellipticals, then one might expect to see a correlation with
apparent ellipticity. J\o{}rgensen \etal\ (1993, 1996) investigated
correlations of ellipticity and $c_4$ parameters with fundamental
plane residuals in an effort to find additional parameters in the FP.
They built realistic models of early-type galaxies and generated
pseudo-observations of them. The data did not, however, follow any of
the trends predicted by the pseudo-observations, despite having taken
all of the observational procedures into account when making the
models.

In Figure \ref{deviate1}(d), we show the residuals as a function of
apparent ellipticity. In the figure, we show the expected deviations
in observed $M/L_V$ ratio for pressure supported E2, E4 and E6
galaxies (solid lines), and for isotropic rotators of intrinsic
flattening 0.6, 0.5, 0.4, 0.3, 0.2, and 0.15 (see, {\it e.g.\/},
\cite{bt87,saglia93}). The data do not follow the predicted curves in
any sensible way. For example the spirals do not behave as flattened
rotators in this diagram, presumably because their residuals are
dominated by stellar population effects (see below).


\subsubsection{Internal Kinematics}

For the spiral galaxies in CL1358+62, Figure \ref{deviate1}(e) shows
that the residuals from the fundamental plane of the early-types is
strongly correlated with central velocity dispersion. This correlation
might suggest that central measurements of the velocity dispersion do
not relate to the virial mass in the same way for the spirals as for
the E/S0s. While it seems reasonable to expect that spiral galaxies
are not homologous with E/S0s, Kelson \etal\ (1999a) showed that the
rotation curves of the two families are similar, at least to radii of
approximately $2r_e$, at the level of a few percent. We therefore
conclude that systematic differences in the central $M/L_V$ ratios of
the two families of galaxies are not arising from structural effects.


\subsubsection{Stellar Populations}

The other possibility is that the properties of the
luminosity-weighted stellar populations are systematically varying as
a function of $\sigma$, at a given $r_e$. Such a trend would suggest
that the correlation in Figure \ref{deviate1}(e), and the low value
for $\alpha$ for the spirals, may actually be rooted in a systematic
variation of stellar populations with $\sigma$. This idea appears to
be borne out by Figure \ref{deviate1}(f), in which deviations from the
fundamental plane are strongly correlated with residual from the
color-magnitude relation, in a manner consistent with simple models of
passive stellar evolution (shown as the dotted lines). The models
overlayed in (f) are GISSEL96 models of single-burst stellar
populations of different metallicities, all normalized to a formation
epoch of $z=2$ ($H_0=65$ \kms\Mpc-1, $q_0=0.1$). Thus, the lines are
progressions of age (time since the burst) from the blue to the red.

Together, the correlations in (e) and (f) would produce a fundamental
plane for the early-type spirals which has a different slope than the
plane of the E/S0s.


\subsubsection{Environment}

Studies of nearby galaxies have shown that position within a cluster is
important for galaxy evolution ({\it e.g.\/} the morphology-density
relation, \cite{dress80,morph2}). Our cluster imaging covers \about
$1.5h^{-1}$ Mpc $\times 1.5 h^{-1}$ Mpc, well-suited for studies of the
morphology-density relation at high redshift (\cite{dgf99}). Within our
HST mosaic, the S0s and spirals do indeed appear more spatially extended
than the ellipticals (\cite{vdcm,dgf99}).

Given this trend and the result that the color-magnitude diagram is
position dependent (\cite{vdcm}), one should expect a correlation
between FP residual and distance from the cluster center. In Figures
\ref{res3}(a,b), we see that there is no {\it obvious\/} trend with
clustercentric radius. The sample is too small and the uncertainties are
simply too large to draw any conclusions with the current sample. More
observations are required to determine the magnitude of any effect. The
environmental dependence of the color-magnitude relation is seen in our
sample in Figures \ref{res3}(c,d) but it is not as obvious as in the
full sample of van Dokkum \etal\ (1998a). As a reminder, those authors
used nearly 200 galaxies in their analysis of the color-magnitude
diagram of the cluster.

Let us now continue with a more in-depth discussion of the stellar
populations.


\section{Modeling the Stellar Populations of the Early-Type Spirals}
\label{models}

In the previous section, we showed that the residuals of the spirals
from the FP and CM relations are predominantly due to their stellar
populations. Do the stellar populations of the spirals evolve
differently than those in early-type galaxies? In \S \ref{es0pops} we
introduced single-burst stellar population models which predicted that
the $M/L_V$ ratios themselves should be correlated with $(B-V)$ color.
In Figure \ref{dbv_es0}(a), we showed that this prediction is
consistent with the data for the early-type galaxies.

In Figure \ref{dbv_all} we expand the plot of $\log M/L_V$ as a
function of galaxy color to include the entire sample. The
least-squares fit to the E/S0s, given in \S \ref{es0pops}, is shown by
the thick solid line. As in the Figure \ref{dbv_es0}(a), the
1-$\sigma$ uncertainties from that fitted slope are indicated by the
shaded region. Single-burst stellar populations with a Salpeter (1955)
IMF are expected to evolve along the thin solid line. In this diagram,
we plot a quadratic approximation to the model because the linear,
analytical expressions in \S \ref{es0pops} were not valid for
extremely young stellar populations. The emission-free early-type
spirals, including the E+A galaxies, are not inconsistent with the
extrapolation of the E/S0s. A least-squares fit to the spirals which
do not show emission or E+A spectra, {\it i.e.\/}, appear to be older
than a few Gyr, yields a slope of $\log M/L_V \propto (3.23\pm 1.23)
(B-V)$, consistent with the slope found using the early-types. With
larger samples and measurements of line strengths, one should be able
to reduce the uncertainties and use such diagnostics to place
constraints on composite stellar populations, or, for example, the
universality of the IMF.

\subsection{Age Effects}

If the fundamental plane and color-magnitude residuals are due to
differences in mean luminosity-weighted ages alone, recall that the
model predicts $\Delta \log M/L_V = (4.52-1.0x)\Delta (B-V)$. If the IMF
follows $x=1.35$ (\cite{imf}), then the models give $\Delta \log M/L_V
\propto 3.1 \Delta (B-V)$. Also recall that the analytical expressions
on which this approximation is based are only valid for stellar
populations with ages greater than 2 Gyr. Thus, in order to test this
prediction, we perform a least-squares fit to the residuals of those
spirals with spectroscopically early-type spectra. This fit, which
excludes the E+A galaxies and the star-forming, emission-line galaxy,
gives $\Delta \log M/L_V \propto (2.32\pm 0.65)\Delta (B-V)$, shown as
the dashed line in Figure \ref{deviate1}(f). This fit implies an IMF
slope of $x=2.0\pm 0.8$.

The spiral galaxies have clearly experienced more recent
star-formation than the ellipticals and S0s. We would now like to
correct for the differences in the ages of their stellar populations.
Since we would like to use the E+A galaxies, which have stellar
populations younger than 2 Gyr, we need a more appropriate analytical
expression relating $\Delta \log M/L_V =f[\Delta (B-V)]$. In fitting a
solar metallicity model, with a Salpeter (1955) IMF, between ages of
750 Myr and 8 Gyr, we find
\begin{equation}
\Delta \log M/L_V = 3.22[\Delta (B-V)]^2+3.26\Delta (B-V)
\label{eq:younger}
\end{equation}
This quadratic is shown as the solid line in Figure \ref{deviate1}(f).
For galaxies with similar radial gradients in their stellar
populations, only $\langle I\rangle_e$ is directly affected by changes
in $M/L_V$ ratio. We therefore correct the surface brightnesses by
\begin{equation}
\langle I\rangle_e^\prime = \langle I\rangle_e\times 10^{3.22[\Delta
(B-V)]^2+3.26\Delta(B-V)}.
\end{equation}
In fitting a new fundamental plane to the corrected spirals, we find
$\alpha'=1.04\pm 0.42$ and $\beta'=-0.61\pm 0.09$, similar to the
J\o{}rgensen \etal\ (1996) values we adopted earlier, though with
larger errors. The scatter about this evolved plane is still \about
30\% in $M/L_V$ ratio, leading to increased uncertainties in the fit.
After correcting for color residuals, no further correlations of
$\Delta \log M/L_V$ with $n$, $BF$, or $\sigma$ are seen.

We conclude that after several Gyr, barring any further
star-formation, or dynamical evolution ({\it i.e.\/}, $r_e$ or
$\sigma$ do not change with time), these cluster spirals will fall on
the fundamental plane of the E/S0s. In Figure \ref{mass2}, we show
that the mass and length scales are comparable to the early-type
sample, indicating that, once evolved, the early-type spirals would
occupy similar locations on the fundamental plane as the E/S0s.

The scatter in the spirals, however, remains quite high compared to
the E/S0s, at about 30\% in $M/L_V$ ratio. This high scatter may be
attributed to a larger spread in possible more complex star-formation
histories, to the greater role of dust in these galaxies, and possibly
to the greater role of rotational support. An important question
remains: does some mechanism transform the spirals into earlier
morphological types while their stellar populations evolve? If such a
transformation occurs, these early-type spirals may be the progenitors
of some nearby S0 galaxies (see \cite{harald}). Given that Dressler
\etal\ (1997) found reduced numbers of S0s and increased numbers of
spirals in $z=0.55$ clusters, perhaps we are witnessing an early stage
in the history and transformation of low-mass S0s.


\subsection{The E+A Galaxies}

For the S0/a and later-type galaxies, $\Delta \log M/L_V$ was roughly
proportional to $\log \sigma$. We interpret this as a correlation of
the mean luminosity-weighted ages of early-type spirals with velocity
dispersion (see \S \ref{residuals}):
\begin{equation}
\Delta \log t \sim (1.22-0.28x)^{-1}\log \sigma.
\end{equation}
Therefore, for single-burst stellar populations, with a Salpeter
(1955) IMF, $\Delta \log t \sim 1.2\log \sigma$. The implication is
that $\sigma=100$ \kms\ early-type spirals in clusters are \about
$1/4$ the luminosity-weighted ages of their $\sigma=300$ \kms\
counterparts. However, if there is a a systematic variation of age
along the fundamental plane of E/S0s, then this interpretation of the
residuals may only be a lower limit on the range of the
luminosity-weighted ages of the cluster spirals.

The E+A galaxies are at the extreme end of this correlation of stellar
populations and $\sigma$. Their $M/L_V$ ratios are about $-0.5$ dex
too low to be consistent with an extrapolation of the fundamental
plane of the ellipticals. Using the Salpeter (1955) IMF, we find that
$\Delta \log t \sim -0.6$, which is equivalent to stating that the
mean luminosity-weighted ages of the E+As in CL1358+62 are about $1/4$
the mean luminosity-weighted age of the cluster ellipticals of the
same $r_e$ and $\sigma$. The implication is that these galaxies
underwent their last burst of star-formation sometime between $z=0.4$
to $0.6$. This constraint is limited by the observational errors, and
the fact that the galaxies likely have composite stellar populations,
though it is consistent with presence of A stars in their spectrum.
The logical extension of the correlation of FP residuals with velocity
dispersion is that the cluster spirals with larger velocity
dispersions, which have FP residuals of $\Delta \log M/L_V \sim -0.3$,
must have undergone their last epochs of star-formation at $z\sim 0.7$
(and may themselves been in an E+A phase at $z\sim 0.5$).

At the present epoch, 3-4 Gyr later, the E+As should be \about $0.08$
mag bluer than the $(B-V)$ CM relation, and have faded \about 1.5 mag.
Two of these galaxies are slightly brighter than $L^*$ (using $M_V^*
\approx -21.5$ mag at $z=0$), with masses slightly lower than $M^*$.
The third has a mass a full dex below $M^*$. One should therefore find
the evolved counterparts of the two massive E+As in the CM relations
of nearby clusters. In Figure 1 of Bower, Lucey, \& Ellis (1992), one
sees large scatter in $(U-V)$ vs. $V_T$ fainter than $L^*$. There is
an abundance of galaxies with $\Delta (U-V)\approx -0.16$ mag ($\Delta
(U-V) \approx 2\Delta (B-V)$). We conclude that the intermediate
redshift star-formation in spirals, even the strong bursts thought to
take place in E+A galaxies (\cite{zabea}), can easily be reconciled
with the present-day samples of nearby early-type cluster galaxies.
While these spiral galaxies may be able to hide in the normal scaling
relations of early-type galaxies, the CL1358+62 E+A galaxies are less
massive than $M^*$ galaxies. We conclude that the E+A galaxies, and
other low velocity dispersion early-type spirals in this CL1358+62
sample, could evolve into present-day S0 galaxies.


\subsection{Metallicity Effects}

We now explore to what extent the observed correlation in Figure
\ref{deviate1}(f) may be metallicity dependent. If the residuals are
due to a systematic trend with $\log Z$, then, using Equations
\ref{bzml} and \ref{bzbv},
\begin{equation}
\Delta \log M/L_V = {{0.56 + 0.15\log Z}\over{0.34+0.054\log Z}}
\times \Delta (B-V)
\end{equation}
For a sample of galaxies whose mean metallicity is solar, then $\Delta
\log M/L_V \propto 1.21 \Delta(B-V)$. This prediction is ruled out by
the least-squares fit to Figure \ref{deviate1}(f) at the 3-$\sigma$
level.

We therefore conclude that it is the presence of young stellar
populations in the spirals which in turn produce a systematic
deviation of the low velocity dispersion spirals from the fundamental
plane of the ellipticals. Future analysis of the metal line strengths
will help determine to what extent the IMF may differ from the one
adopted in the models, and to what extent metallicity variations are
present in our sample.

In Equation \ref{eq:age} we showed the correlation expected between FP
and CM residuals for old stellar populations, when variations in mean
luminosity-weighted age are the primary cause of the residuals. Recall
that the observed correlation for the spirals was $\Delta \log M/L_V
\propto (2.5\pm 0.8) \Delta (B-V)$. Using this correlation, the
single-burst models imply
\begin{equation}
\Delta \log Z/ \Delta \log t =
(-0.67_{-0.28}^{+0.16}) x + (1.05_{+0.90}^{-0.50})
\end{equation}
An IMF with $x=1.35$ implies $\Delta \log Z/ \Delta \log t =
0.14_{-0.28}^{+0.52}$, ruling out the Worthey (1994) "$3/2$" rule
by nearly 3-$\sigma$. If this "3/2" rule is true, then the slope of
the IMF (still parameterized as a single power-law) must be $x\sim
2.5$. Analysis of the line-strengths is required to fully understand
any correlations between luminosity-weighted age and metallicity, as
well as conclusively test for variations in the initial mass function.

Though the spirals' FP is tilted with respect to the plane of the
ellipticals, this can be accounted for using stellar population
models. However, the residuals about the corrected plane are still
quite large. This remaining scatter could be due to additional random
variations in the stellar populations, such as from metallicity, or
other unknown factors such as galaxy shapes, or even variable dust
content.


\section{Conclusions}
\label{conclusions}

The fundamental plane in the cluster CL1358+62 has been measured using
accurate internal kinematics and structural parameters. In total, we
have 53 cluster members, three of which have E+A spectra, and one
which has emission lines, indicative of star-formation. Of the 53, 31
have been classified as E, E/S0, or S0 galaxies. There are 13 S0/a
galaxies, 6 with Sa morphology, two Sab galaxies, and 1 Sb. Because
our sample was selected without the use of morphology, we have made a
thorough analysis of the cluster population as a whole. Several
conclusions can be drawn from the current data.

(1) \ \ The 30 E, E/S0, and S0s which do not have E+A spectra or strong
emission lines follow a relation which is similar in form to that
found in nearby clusters:
\begin{equation}
\log r_e
\propto (1.31\pm 0.13)\log\sigma - (0.86\pm 0.10)\log\langle I\rangle_e
\end{equation}
Using these slopes, the ellipticals and S0s give identical fundamental
plane zero-points, similar to the ellipticals and S0s in nearby
clusters (\cite{jfk96}). Fitting individual relations to the visually
classified ellipticals and S0s does not produce significantly
different results.

(2) \ \ The $M/L_V$ ratios of the CL1358+62 early-types are lower than
those in Coma by $-0.13\pm 0.03$ dex ($q_0=0.1$, $-0.10\pm 0.03$ dex
for $q_0=0.5$). The look-back time to CL1358+62, about a third the
present age of the Universe, is therefore equivalent to about $1/4$ of
the lifespan of nearby cluster galaxies, in agreement with a high mean
redshift of formation for the stars in E/S0s. This comparison assumes
that these Es and S0s in CL1358+62 are co-eval with those in Coma. If
any contemporary E/S0 galaxies possessed later-type morphologies at
intermediate redshifts, they would not have been included in the
determination of this offset. Therefore, the measured $M/L_V$ offset
may only reflect a subset of contemporary E/S0 galaxies
(\cite{franx95}).

(3) \ \ The stellar populations of early-type galaxies vary along the
fundamental plane such that $\log M/L_V \propto (2.55 \pm 0.79) \times
(B-V)$. The current data are not sufficient to specify to what extent
this correlation derives from age and/or metallicity variations. If
one corrects the surface brightnesses of the E/S0s for the observed
correlation, the resulting plane agrees with the expectation predicted
by the virial theorem and homology ($r\propto \sigma^2 I^{-1}$). This
result suggests that early-type galaxies form an homologous family.

(4) \ \ The E/S0 galaxies have an intrinsic 1-$\sigma$ scatter of 14\%
in $V$-band $M/L$ ratio about the local FP slope. This scatter is
consistent with the scatter in the color-magnitude relation
(\cite{vdcm}), assuming that it is caused by a scatter in the
luminosity-weighted ages of the stellar populations. The implication
is that the CL1358+62 elliptical galaxies would have a scatter in mean
luminosity-weighted ages of \about 15\%, depending on the shape of the
IMF. The implication is that nearly all of the star-formation in the
CL1358+62 E/S0s was finished by $z=1$. Measurements of Balmer and
metal line strengths are required to fully understand to what extent
random variations in age and metal abundance are causing scatter in
the two scaling relations.

(5) \ \ The early-type spirals follow a different fundamental plane
relation from the ellipticals and S0s. Given this, we studied the
residuals of the spirals with respect to the fundamental plane of the
ellipticals and find that the residuals of the spirals correlate with
both velocity dispersion and residual from the color-magnitude
relation (\cite{vdcm}). Thus, the spirals follow a different plane
from the ellipticals and S0s because of a systematic variation of
stellar populations with velocity dispersion. After several Gyr of
passive stellar evolution, these spirals will follow a plane with the
same form as the E/S0s, but with scatter that is twice as large.

(6) \ \ The smooth transition of color and FP residuals towards older
(redder, higher $M/L_V$ ratio) spirals indicates that star-formation
has been a continuous process for these galaxies. The galaxies with
the most recent star-formation have been intermediate and low velocity
dispersion early-type spirals. The systematic correlation of FP
residual with color, and, for example, the strong Balmer absorption
lines, indicate that cluster spirals have been forming stars up
through times as late as $z\sim 0.5$. The spiral galaxies contain
composite stellar populations, and their long-term star-formation
histories cannot be modeled adequately with the current data.
Regardless, these galaxies clearly have a broad dispersion in
properties, unlike the E/S0 galaxies.

(7) \ \ The correlation of FP residual and color residual for
early-type spirals extends to the E+A galaxies, forming a tail in a
sequence of recent, or recently extinguished, star-formation. That the
E+A and emission line galaxies are very young and have low velocity
dispersions simply places them at the extreme end of a relation
between stellar populations and velocity dispersion which is specific
to the spiral galaxies. Simple stellar population models imply that
the mean luminosity-weighted ages of these spiral galaxies follows
$\Delta \log t \sim (1.22-0.28x)^{-1}\log \sigma.$ Thus, an Sa galaxy
with $\sigma=100$ \kms\ would be about $1/4$ the mean
luminosity-weighted age of an Sa with $\sigma= 300$ \kms, for a
Salpeter (1955) IMF. However, this interpretation depends upon
whatever systematic variations of luminosity-weighted ages may exist
along the fundamental plane of early-type galaxies.

\begin{figure}
\centerline{\mkfigbox{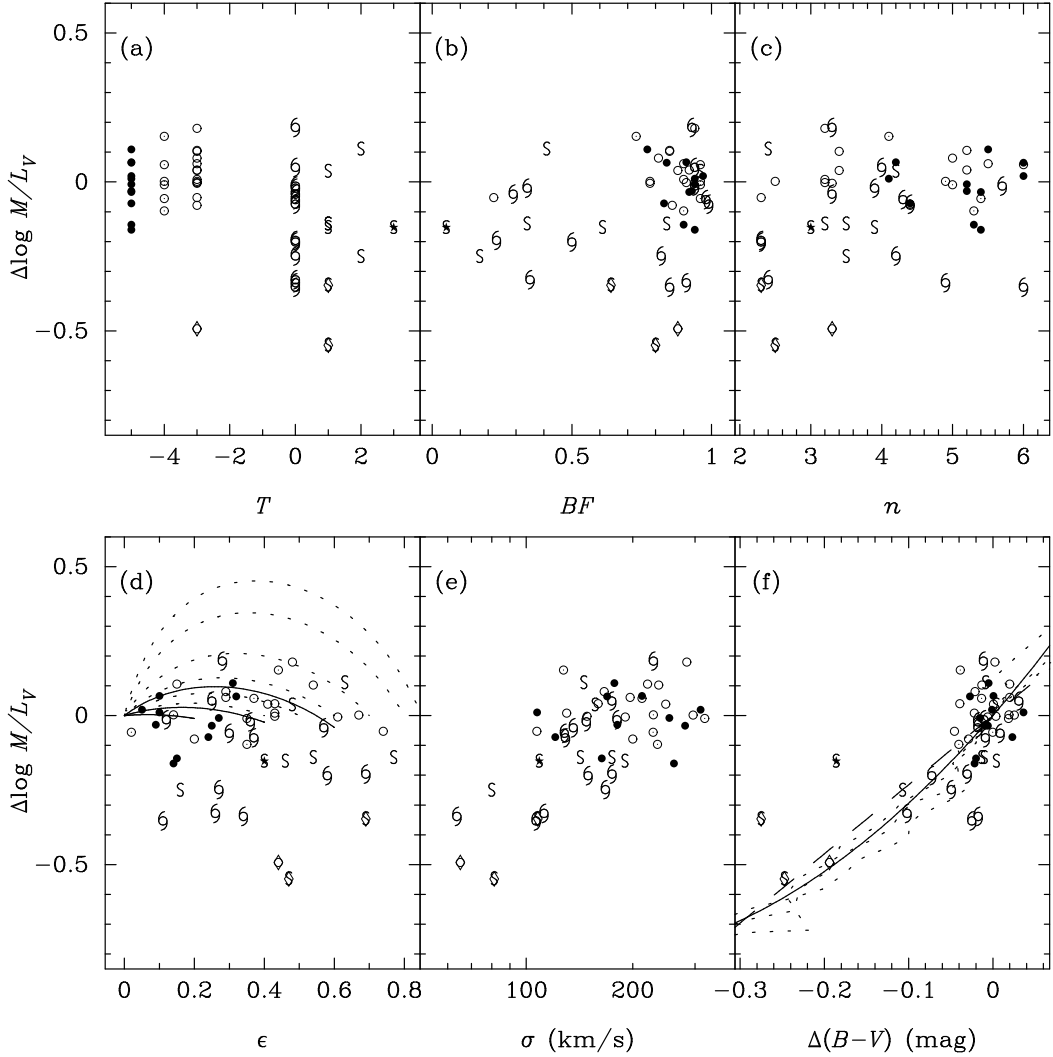}{0.75\textwidth}}
\figcaption[fig5.ps]{
Fundamental plane residual as a function of (a) $T$, galaxy
morphology, as determined in Fabricant \etal\ (1998); (b) bulge
fraction, derived from bulge-plus-disk decompositions of the
integrated surface brightness profiles; (c) $n$, the shape parameter
describing the best-fit $r^{1/n}$-law; (d) apparent ellipticity
superimposed with curves following the random projection of simple
models described in the text; (e) central velocity dispersion; and (f)
residual from the color-magnitude relation (\cite{vdcm}), superimposed
with models of single-burst stellar populations with three different
metallicities ($\rm [Z/H]=-0.4, 0.0, +0.4$). The models, normalized to
a redshift of formation of $z=2$ ($H_0=65$ \kms\Mpc-1, $q_0=0.1$) at
zero residual, are shown as dotted lines. The least-squares fit to the
spirals without emission or strong Balmer absorption of $\Delta \log
M/L_V \propto 2.5 \Delta (B-V)$ is shown as the dashed line. The
analytical approximation to a solar metallicity, $x=1.35$,
single-burst stellar population is shown as the solid line. There is a
clear trend of $M/L_V$ residual with galaxy morphology, central
velocity dispersion, and color. The symbols are as in Figures
\ref{view1} and \ref{view2}.
\label{deviate1}}
\end{figure}

\begin{figure}
\centerline{\mkfigbox{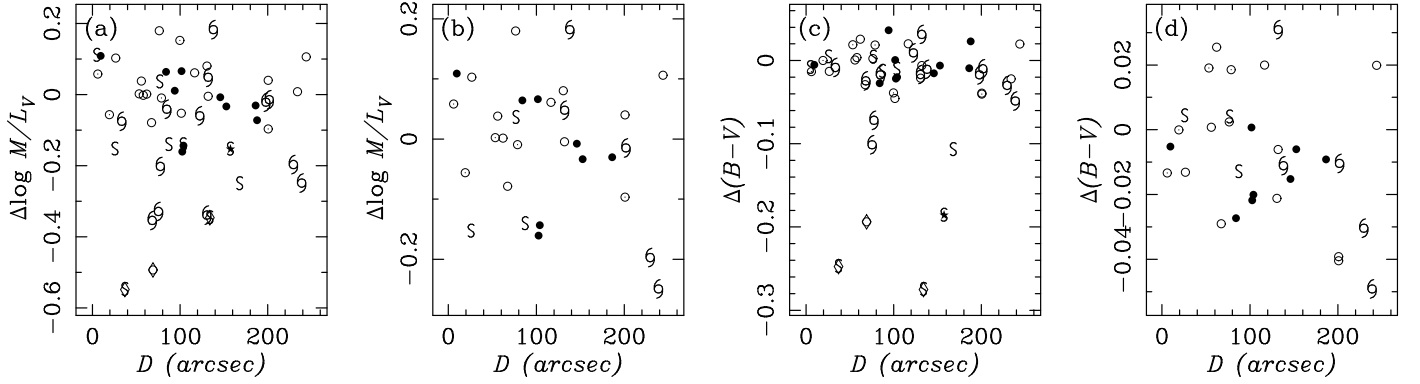}{\textwidth}}
\figcaption[fig6.ps]{
Fundamental plane residuals plotted vs. distance from the cD for (a) the
full sample, and (b) for those galaxies with $\sigma \ge 150$ \kms. In (c)
and (d), the color-magnitude relation residuals are shown in the same
manner. At the distance of the cluster, the scale is 4.8 kpc/arcsec. No
significant correlation of fundamental plane residual with distance
from the cD is seen.
\label{res3}}
\end{figure}

\begin{figure}
\centerline{\mkfigbox{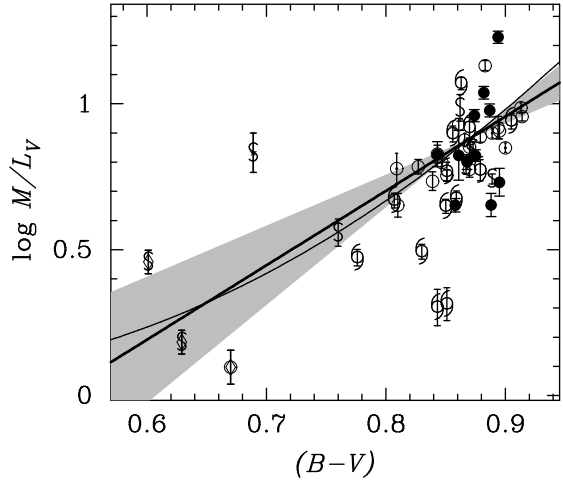}{0.4\textwidth}}
\figcaption[fig7.ps]{
$M/L_V$ ratios are shown plotted against $(B-V)$ colors for the
full sample. The linear-least squares fit from \S \ref{es0pops} is
shown by the thick solid line and enclosed gray region. For a
Salpeter (1955) IMF, and solar metallicities, one expects the $M/L_V$
ratio and $(B-V)$ colors to evolve along paths parallel to the thin
solid line.
\label{dbv_all}}
\end{figure}

\begin{figure}
\centerline{ \mkfigbox{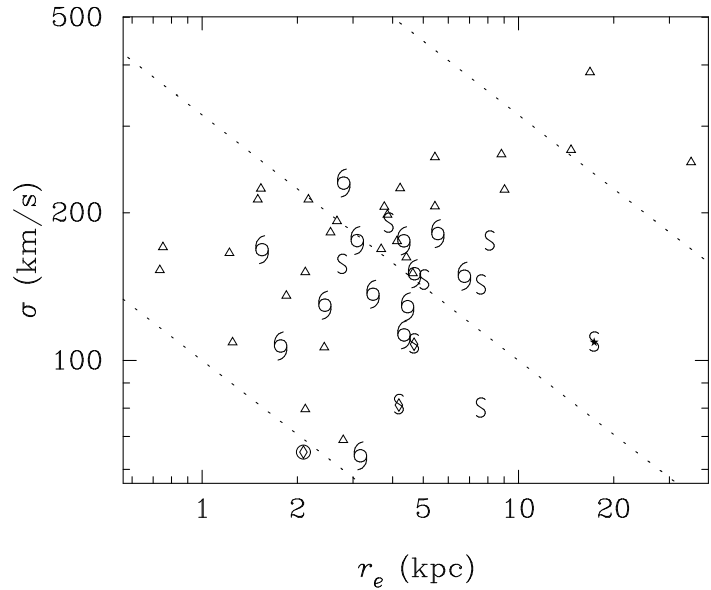}{0.5\textwidth}}
\figcaption[fig8.ps]{
Similar to the E/S0s shown in Figure \ref{mass}, the later-type
galaxies in our sample also have mass and length scales which are
comparable to those seen in the local sample. As in Fig. \ref{mass},
contours of constant mass ($5\sigma^2r_e/G=10^{10,11,12}M_\odot$) are
shown by the dotted lines.
\label{mass2}}
\end{figure}

(8) \ \ Early-type spirals brighter than $L^*$ at the present epoch
must have formed before $z>1$. Furthermore, the implication of Figure
\ref{deviate1}(e,f) is that early-type spirals with $\sigma \gg 100$
\kms\ should show evidence for more recent star-formation at higher
redshifts. This has been tentatively seen in MS1054--03 at $z=0.83$
(see \cite{vdfp98}).

(9) \ \ The E+A galaxies could be as much as three times too bright
for their $r_e$ and $\sigma$, compared to an extrapolation of the
fundamental plane of the ellipticals. The E+As are consistent with
having a burst of star-formation at $z\sim 0.5\pm 0.10$. These data do
not place any stronger constraints due to observational errors, and
the unknown extent to which the stellar populations are composites of
both old and young stars. The three E+A galaxies in the sample span a
range of masses, the highest nearly the mass of an $L^*$ galaxy, down
to a full dex below $M^*$. By $z\approx 0$, these galaxies will have
faded about 1.5 mag and only be $\Delta (B-V)\approx 0.08$ mag bluer
than local color-magnitude relations ({\it i.e.\/}, \cite{bower}),
assuming these galaxies can be adequately modeled by a single,
instantaneous burst. The E+As have variable bulge fractions, but
invariably have disks and are partly supported by rotation
(\cite{kelson99a},b, \cite{wirth,franx93}). It is unlikely that they
will become present-day ellipticals, but, like the other young spirals
in the sample, they may evolve into S0s.

The fundamental plane has become a useful tool for studying the
histories of galaxies later than E and S0 (cf. \cite{bbfn}). The
larger spiral fractions in clusters at higher redshift will allow for
a better understanding of the evolution and transition of the spiral
and early-type populations ({\it e.g.\/}, \cite{vdfp98}). If there are
processes which turn spiral galaxies into S0s between $z=0.33$ and
$z=0$, then such processes must somehow work to reduce their scatter
about the fundamental plane, unless such galaxies were excluded from
nearby samples for reasons other than morphology.

Future work will be aimed at measuring absorption line strengths and
near-IR colors. Together with the $V$-band $M/L$ ratios, one will be
able to expand the modeling with the goal of disentangling composite
stellar populations ({\it e.g.\/}, in the E+A galaxies),
metallicities, dust contamination, the IMF, and galaxy shapes.
Analysis of the fundamental plane of field galaxies at intermediate
redshifts will be a very important step for comparing the evolutionary
histories of early-type galaxies in different environments. Analysis
of the early-types in poor groups and the field will also be important
in relating the star-formation histories in intermediate environments
to those in clusters and the field.

Barring the transformation of spirals into early-types, present-day
cluster Es and S0s are among the oldest objects in the universe, and
have undergone little, if any, late-time star formation.

\acknowledgements

We gratefully acknowledge D. Koo and S. Faber, who provided valuable
comments on an early version of the paper. We also would like to
acknowledge referee, G. Oemler, who helped to strengthen the
presentation of this work. Furthermore, we appreciate the effort of
all those in the HST program that made this unique Observatory work as
well as it does. The assistance of those at STScI who helped with the
acquisition of the HST data is also gratefully acknowledged. We also
appreciate the effort of those at the W.M.Keck observatory who
developed and supported the facility and the instruments that made
this program possible. Support from STScI grants GO05989.01-94A,
GO05991.01-94A, and AR05798.01-94A, and NSF grant AST-9529098 is
gratefully acknowledged.


\newpage

\clearpage

\clearpage

\clearpage

\clearpage

\clearpage

\begin{deluxetable}{r l r r r r c c c r l r r r r c c}
\tiny
\tablewidth{0pt}
\tablecaption{Data used in the CL1358+62 Fundamental
Plane\label{table}}
\tablehead{
\colhead{ID} &
\colhead{Type} &
\colhead{$n$} &
\colhead{$\sigma$} &
\colhead{$r_e$} &
\colhead{$\langle \mu_e\rangle $} &
\colhead{$R$} &
\colhead{$(B-V)_z$} &&
\colhead{ID} &
\colhead{Type} &
\colhead{$n$} &
\colhead{$\sigma$} &
\colhead{$r_e$} &
\colhead{$\langle \mu_e\rangle $} &
\colhead{$R$} &
\colhead{$(B-V)_z$}}
\startdata
095&S0  & 5.2& 220.5&  0.356& 19.84& 20.18& 0.894&&                                  359&S0  & 4.4& 200.3&  0.475& 19.95& 20.14& 0.851\nl                                 
\nodata&\nodata&$\pm 1.2$&$\pm   6.4$&$\pm  0.004$&$\pm  0.02$&\nodata&\nodata&& \nodata&\nodata&$\pm 0.4$&$\pm   4.5$&$\pm  0.005$&$\pm  0.02$&\nodata&\nodata\nl
110&S0/a& 2.3& 174.8&  0.648& 20.32& 19.87& 0.850&&                                  360&E   & 5.5& 177.5&  0.241& 19.65& 20.74& 0.861\nl                                 
\nodata&\nodata&$\pm 0.3$&$\pm   3.7$&$\pm  0.012$&$\pm  0.04$&\nodata&\nodata&& \nodata&\nodata&$\pm 0.5$&$\pm  19.7$&$\pm  0.004$&$\pm  0.03$&\nodata&\nodata\nl
129&S0/a& 4.2& 167.5&  0.324& 19.27& 19.94& 0.830&&                                  366&S0/a& 3.3& 228.9&  0.587& 20.68& 20.31& 0.863\nl                                 
\nodata&\nodata&$\pm 0.2$&$\pm   3.8$&$\pm  0.004$&$\pm  0.02$&\nodata&\nodata&& \nodata&\nodata&$\pm 0.3$&$\pm   6.4$&$\pm  0.008$&$\pm  0.03$&\nodata&\nodata\nl
135&S0  & 5.2& 159.7&  0.351& 20.19& 20.59& 0.827&&                                  368&Sab & 2.4& 145.6&  1.055& 22.09& 20.33& 0.862\nl                                 
\nodata&\nodata&$\pm 0.2$&$\pm   5.1$&$\pm  0.009$&$\pm  0.05$&\nodata&\nodata&& \nodata&\nodata&$\pm 2.6$&$\pm   7.5$&$\pm  0.032$&$\pm  0.06$&\nodata&\nodata\nl
142&S0/a& 3.9& 147.9&  1.414& 22.15& 20.16& 0.856&&                                  369&S0/a& 4.3& 128.2&  0.935& 21.70& 20.51& 0.879\nl                                 
\nodata&\nodata&$\pm 0.9$&$\pm   4.4$&$\pm  0.046$&$\pm  0.07$&\nodata&\nodata&& \nodata&\nodata&$\pm 0.3$&$\pm   6.1$&$\pm  0.023$&$\pm  0.05$&\nodata&\nodata\nl
164&S0/a& 5.7& 180.6&  1.165& 21.56& 19.83& 0.870&&                                  371&Sa  & 3.5& 174.8&  1.699& 21.84& 20.02& 0.871\nl                                 
\nodata&\nodata&$\pm 0.7$&$\pm   4.3$&$\pm  0.035$&$\pm  0.06$&\nodata&\nodata&& \nodata&\nodata&$\pm 1.5$&$\pm   4.0$&$\pm  0.033$&$\pm  0.04$&\nodata&\nodata\nl
182&S0  & 3.2& 130.1&  0.418& 20.70& 20.82& 0.839&&                                  372&Sa  & 3.2& 142.3&  1.595& 22.09& 20.02& 0.867\nl                                 
\nodata&\nodata&$\pm 0.2$&$\pm   5.4$&$\pm  0.008$&$\pm  0.04$&\nodata&\nodata&& \nodata&\nodata&$\pm 1.2$&$\pm   5.0$&$\pm  0.064$&$\pm  0.08$&\nodata&\nodata\nl
209&Sa  & 2.3& 107.9&  0.980& 21.34& 19.99& 0.601&&                                  375&E   & 6.0& 311.3&  3.813& 22.44& 19.09& 0.894\nl                                 
\nodata&\nodata&$\pm 0.3$&$\pm   5.2$&$\pm  0.018$&$\pm  0.04$&\nodata&\nodata&& \nodata&\nodata&$\pm 0.1$&$\pm   7.0$&$\pm  0.075$&$\pm  0.04$&\nodata&\nodata\nl
211&S0  & 3.3& 190.5&  0.612& 20.58& 20.14& 0.870&&                                  381&E/S0& 6.0& 212.4&  0.392& 19.92& 19.99& 0.866\nl                                 
\nodata&\nodata&$\pm 0.3$&$\pm   6.4$&$\pm  0.015$&$\pm  0.05$&\nodata&\nodata&& \nodata&\nodata&$\pm 0.1$&$\pm   5.5$&$\pm  0.008$&$\pm  0.04$&\nodata&\nodata\nl
212&E   & 5.2& 181.0&  0.547& 20.44& 20.08& 0.868&&                                  397&S0/a& 3.3& 135.9&  0.727& 21.29& 20.52& 0.851\nl                                 
\nodata&\nodata&$\pm 0.2$&$\pm   5.7$&$\pm  0.009$&$\pm  0.03$&\nodata&\nodata&& \nodata&\nodata&$\pm 0.3$&$\pm   5.1$&$\pm  0.020$&$\pm  0.06$&\nodata&\nodata\nl
215&S0  & 5.0& 166.2&  0.338& 20.17& 20.74& 0.843&&                                  408&S0  & 3.4& 237.1&  0.353& 19.70& 20.20& 0.861\nl                                 
\nodata&\nodata&$\pm 1.0$&$\pm   5.9$&$\pm  0.006$&$\pm  0.03$&\nodata&\nodata&& \nodata&\nodata&$\pm 0.4$&$\pm   6.2$&$\pm  0.003$&$\pm  0.02$&\nodata&\nodata\nl
233&E/S0& 5.3& 234.9&  0.590& 19.95& 19.36& 0.847&&                                  409&E   & 4.1& 107.3&  0.536& 21.38& 21.17& 0.895\nl                                 
\nodata&\nodata&$\pm 0.3$&$\pm   4.5$&$\pm  0.012$&$\pm  0.04$&\nodata&\nodata&& \nodata&\nodata&$\pm 1.1$&$\pm   6.7$&$\pm  0.005$&$\pm  0.02$&\nodata&\nodata\nl
234&Sb  & 3.0& 108.9&  3.636& 23.68& 20.08& 0.689&&                                  410&S0  & 3.2& 148.6&  0.497& 20.70& 20.78& 0.870\nl                                 
\nodata&\nodata&$\pm 1.0$&$\pm   8.2$&$\pm  0.349$&$\pm  0.18$&\nodata&\nodata&& \nodata&\nodata&$\pm 0.2$&$\pm   5.2$&$\pm  0.008$&$\pm  0.03$&\nodata&\nodata\nl
236&S0  & 5.5& 196.7&  0.539& 20.51& 20.20& 0.895&&                                  412&E   & 6.0& 169.4&  0.368& 20.22& 20.48& 0.843\nl                                 
\nodata&\nodata&$\pm 1.5$&$\pm   5.6$&$\pm  0.009$&$\pm  0.03$&\nodata&\nodata&& \nodata&\nodata&$\pm 0.1$&$\pm   6.3$&$\pm  0.007$&$\pm  0.04$&\nodata&\nodata\nl
242&E   & 4.2& 212.4&  1.201& 21.54& 19.96& 0.882&&                                  433&S0/a& 6.0& 106.6&  0.371& 19.95& 20.21& 0.851\nl                                 
\nodata&\nodata&$\pm 0.2$&$\pm   5.3$&$\pm  0.026$&$\pm  0.04$&\nodata&\nodata&& \nodata&\nodata&$\pm 0.1$&$\pm   4.0$&$\pm  0.011$&$\pm  0.06$&\nodata&\nodata\nl
256&E   & 5.4& 261.8&  0.956& 20.30& 18.96& 0.875&&                                  440&S0/a& 4.9&  63.7&  0.663& 21.67& 20.96& 0.843\nl                                 
\nodata&\nodata&$\pm 0.4$&$\pm   4.3$&$\pm  0.010$&$\pm  0.02$&\nodata&\nodata&& \nodata&\nodata&$\pm 0.9$&$\pm   5.0$&$\pm  0.016$&$\pm  0.05$&\nodata&\nodata\nl
269&E/S0& 5.0& 319.5&  0.776& 20.05& 19.12& 0.913&&                                  454&S0/a& 2.3& 149.6&  0.983& 21.16& 19.96& 0.807\nl                                 
\nodata&\nodata&$\pm 1.0$&$\pm   5.8$&$\pm  0.008$&$\pm  0.02$&\nodata&\nodata&& \nodata&\nodata&$\pm 0.3$&$\pm   3.7$&$\pm  0.014$&$\pm  0.03$&\nodata&\nodata\nl
292&S0/a& 2.4& 112.3&  0.911& 21.21& 20.14& 0.776&&                                  463&S0  & 3.2& 284.4&  0.667& 20.50& 19.99& 0.883\nl                                 
\nodata&\nodata&$\pm 0.4$&$\pm   3.2$&$\pm  0.031$&$\pm  0.07$&\nodata&\nodata&& \nodata&\nodata&$\pm 0.2$&$\pm   6.4$&$\pm  0.007$&$\pm  0.02$&\nodata&\nodata\nl
298&S0  & 2.5& 296.6&  0.642& 19.93& 19.47& 0.914&&                                  465&Sa  & 4.2& 156.7&  0.581& 20.93& 20.61& 0.873\nl                                 
\nodata&\nodata&$\pm 0.5$&$\pm   5.2$&$\pm  0.007$&$\pm  0.02$&\nodata&\nodata&& \nodata&\nodata&$\pm 0.2$&$\pm   5.0$&$\pm  0.009$&$\pm  0.03$&\nodata&\nodata\nl
300&S0  & 3.4& 248.2&  0.386& 19.59& 19.98& 0.879&&                                  481&S0  & 2.3& 107.2&  0.478& 21.06& 21.07& 0.810\nl                                 
\nodata&\nodata&$\pm 0.4$&$\pm   5.5$&$\pm  0.004$&$\pm  0.02$&\nodata&\nodata&& \nodata&\nodata&$\pm 0.3$&$\pm   5.5$&$\pm  0.015$&$\pm  0.06$&\nodata&\nodata\nl
303&E   & 5.3& 163.5&  0.494& 20.18& 19.95& 0.858&&                                  493&E/S0& 4.1& 127.5&  0.207& 20.09& 21.59& 0.809\nl                                 
\nodata&\nodata&$\pm 0.3$&$\pm   4.3$&$\pm  0.007$&$\pm  0.03$&\nodata&\nodata&& \nodata&\nodata&$\pm 1.1$&$\pm   8.8$&$\pm  0.010$&$\pm  0.10$&\nodata&\nodata\nl
309&E/S0& 4.9& 228.7&  0.433& 19.80& 19.84& 0.900&&                                  523&S0/a& 4.0& 174.6&  0.910& 21.42& 20.36& 0.905\nl                                 
\nodata&\nodata&$\pm 0.9$&$\pm   4.7$&$\pm  0.003$&$\pm  0.01$&\nodata&\nodata&& \nodata&\nodata&$\pm 1.0$&$\pm   5.9$&$\pm  0.016$&$\pm  0.04$&\nodata&\nodata\nl
328&Sa  & 2.5&  81.2&  0.878& 21.15& 19.96& 0.629&&                                  531&E   & 5.4& 281.3&  1.048& 20.63& 19.33& 0.887\nl                                 
\nodata&\nodata&$\pm 0.5$&$\pm   4.2$&$\pm  0.016$&$\pm  0.04$&\nodata&\nodata&& \nodata&\nodata&$\pm 0.4$&$\pm   5.0$&$\pm  0.009$&$\pm  0.02$&\nodata&\nodata\nl
335&S0/a& 4.4& 129.1&  0.512& 20.79& 20.47& 0.859&&                                  534&E   & 4.4& 120.9&  0.464& 20.77& 20.73& 0.888\nl                                 
\nodata&\nodata&$\pm 0.4$&$\pm   4.3$&$\pm  0.009$&$\pm  0.04$&\nodata&\nodata&& \nodata&\nodata&$\pm 0.4$&$\pm   6.1$&$\pm  0.008$&$\pm  0.04$&\nodata&\nodata\nl
343&S0  & 3.3&  65.1&  0.438& 20.66& 20.66& 0.670&&                                  536&E   & 5.2& 254.1&  0.890& 20.63& 19.70& 0.874\nl                                 
\nodata&\nodata&$\pm 0.3$&$\pm   5.0$&$\pm  0.010$&$\pm  0.05$&\nodata&\nodata&& \nodata&\nodata&$\pm 0.2$&$\pm   5.0$&$\pm  0.009$&$\pm  0.02$&\nodata&\nodata\nl
353&E/S0& 5.4& 228.5&  0.958& 20.79& 19.50& 0.889&&                                  549&Sab & 3.5&  79.9&  1.592& 22.77& 20.66& 0.760\nl                                 
\nodata&\nodata&$\pm 0.4$&$\pm   4.4$&$\pm  0.007$&$\pm  0.01$&\nodata&\nodata&& \nodata&\nodata&$\pm 0.5$&$\pm   4.6$&$\pm  0.058$&$\pm  0.07$&\nodata&\nodata\nl
356&Sa  & 3.9& 190.2&  0.815& 20.62& 19.66& 0.889\nl
\nodata&\nodata&$\pm 0.1$&$\pm   3.7$&$\pm  0.011$&$\pm  0.03$&\nodata&\nodata\nl
\enddata
\tablecomments{
Velocity dispersions from Kelson \etal\ (1999a). The effective radii
are given in arcsec. Surface brightnesses are in $V_z$ magnitudes per
square arcsec, where $V_z$ is equivalent to the Johnson $V$ filter
redshifted to the observed frame of the galaxies (Kelson \etal\
1999b). The formal uncertainties in these structural parameters are
reported above, and, although the systematic errors in either
parameter can be quite large, their combined error in the fundamental
plane is quite small. Morphologies are from Fabricant \etal\ (1999),
and $(B-V)_z$ colors are from van Dokkum \etal\ (1998a). The colors are
likely to be uncertain at a level of $\pm 0.02$ mag. The $R$
magnitudes are from Fabricant, McClintock, \& Bautz (1991).}
\end{deluxetable}

\end{document}